\documentclass[pre,twocolumn,amsmath,amssymb,superscriptaddress,showpacs]{revtex4}
\usepackage{fixmath,graphicx,bm,macros,psfrag,array,color,empheq}

\begin{document}

\title{A freely relaxing polymer remembers how it was straightened}

\author{Benedikt Obermayer} \affiliation{Arnold Sommerfeld Center and
  Center of NanoScience, Ludwig-Maximilians-Universit\"at M\"unchen,
  Theresienstr.~37, 80333 M\"unchen} \author{Wolfram M\"obius}
\affiliation{Arnold Sommerfeld Center and Center of NanoScience,
  Ludwig-Maximilians-Universit\"at M\"unchen, Theresienstr.~37, 80333
  M\"unchen} \affiliation{Institut f\"ur Theoretische Physik,
  Universit\"at zu K\"oln, Z\"ulpicher Str.~77, 50937 K\"oln}
\author{Oskar Hallatschek} \affiliation{Max Planck Institute for
  Dynamics and Self-Organization, 37073 G\"ottingen}
 \author{Erwin Frey}\email{frey@physik.lmu.de}
\affiliation{Arnold Sommerfeld Center and Center of NanoScience,
  Ludwig-Maximilians-Universit\"at M\"unchen, Theresienstr.~37, 80333
  M\"unchen} \author{Klaus Kroy}\email{kroy@itp.uni-leipzig.de}
\affiliation{Institut f\"ur Theoretische Physik, Universit\"at
  Leipzig, Postfach 100920, 04009 Leipzig}

\begin{abstract}
  The relaxation of initially straight semiflexible polymers has been
  discussed mainly with respect to the longest relaxation time. The
  biologically relevant non-equilibrium dynamics on shorter times is
  comparatively poorly understood, partly because ``initially
  straight'' can be realized in manifold ways. Combining Brownian
  dynamics simulations and systematic theory, we demonstrate how
  different experimental preparations give rise to specific short-time
  and universal long-time dynamics. We also discuss boundary effects
  and the onset of the stretch--coil transition.  
\end{abstract}

\maketitle

The mechanical properties of semiflexible polymers, which form an
integral part of the cell structure, are of high relevance to the
understanding of cell elasticity and
motility~\cite{bausch-kroy:06,maher:07}.  External stress not only
remarkably changes static and dynamic
features~\cite{gardel-etal:06,fernandez-pullarkat-ott:06,semmrich-etal:07},
but has also important biological implications, e.g., for enzyme
activity on
DNA~\cite{wuite-etal:00,goel-astumian-herschbach:03,vdbroek-noom-wuite:05}.
Particularly intruiging aspects of stress-controlled behavior can be
observed for the relaxation of semiflexible filaments from initially
nearly straight (i.e., highly stressed) conformations. In recent
years, many experimental and theoretical studies have addressed this
paradigmatic problem of polymer rheology~(see, e.g.,
Refs.~\cite{perkins-quake-smith-chu:94,brochard:95,manneville-etal:96,sheng-lai-tsao:97,bakajin-etal:98,brochard-buguin-degennes:99,hatfield-quake:99,ladoux-doyle:00,maier-seifert-raedler:02,turner-cabodi-craighead:02,schroeder-babcock-shaqfeh-chu:03,bohbot_raviv-etal:04,dimitrakopoulos:04,schroeder-shaqfeh-chu:04,reccius-etal:05,shaqfeh:05,wang-gao:05,goshen-etal:05,crut-etal:07,hoffmann-shaqfeh:07}),
often primarily focused on the influence of hydrodynamic interactions
on the longest relaxation time $\tR$, which is a key identifier of the
stretch--coil
transition. 
However, on much shorter times the polymer dynamics is predominantly
controlled by the highly nontrivial \emph{internal} conformational
relaxation~\cite{frey-etal:97,legoff-hallatschek-frey-amblard:02},
which plays a relevant role in many biological situations ranging from
the viscoelastic response of polymer networks~\cite{morse:98c} to
molecular motor kinetics~\cite{legoff-amblard-furst:02} and DNA
supercoiling dynamics~\cite{crut-etal:07, koster-etal:07}. This aspect
of the relaxation is still poorly understood, the more so as standard
analytical techniques based on linearized equations of motion fail due
to inherent nonlinearities initiated by strong
perturbations~\cite{seifert-wintz-nelson:96,hallatschek-frey-kroy:05}.
Further, because a \emph{completely} straightened polymer conformation
can in practice not be realized in the presence of thermal noise from
the environment, the short-time dynamics of an initially ``nearly''
straight filament will reflect the way it was straightened: filaments
can be stretched by optical
tweezers~\cite{bohbot_raviv-etal:04,goshen-etal:05}, by electric
fields~\cite{bakajin-etal:98,maier-seifert-raedler:02,turner-cabodi-craighead:02,reccius-etal:05,balducci-hsieh-doyle:07},
or by flows of different
geometry~\cite{perkins-quake-smith-chu:94,manneville-etal:96,perkins-smith-chu:97,bakajin-etal:98,ladoux-doyle:00,schroeder-babcock-shaqfeh-chu:03,schroeder-shaqfeh-chu:04,hoffmann-shaqfeh:07},
but a straightened contour can also result from low initial
temperatures.  In any case the relaxation dynamics is driven
exclusively by stochastic forces. This raises the question how results
obtained with different setups should be compared and when the
dependence on initial conditions fades out.

\begin{figure}
  \includegraphics[width=.23\textwidth]{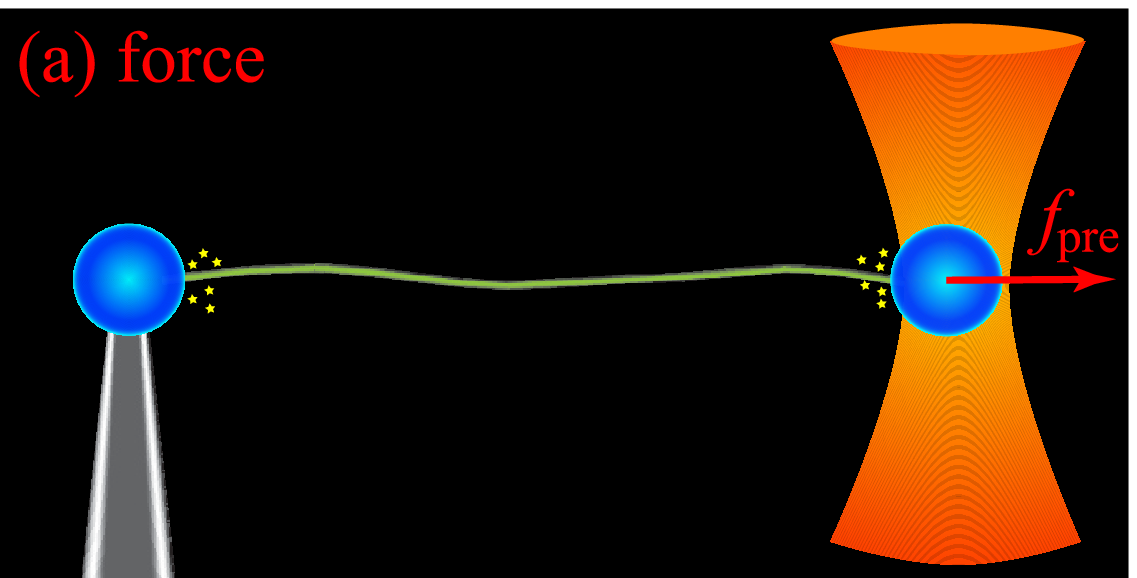}\hfill
  \includegraphics[width=.23\textwidth]{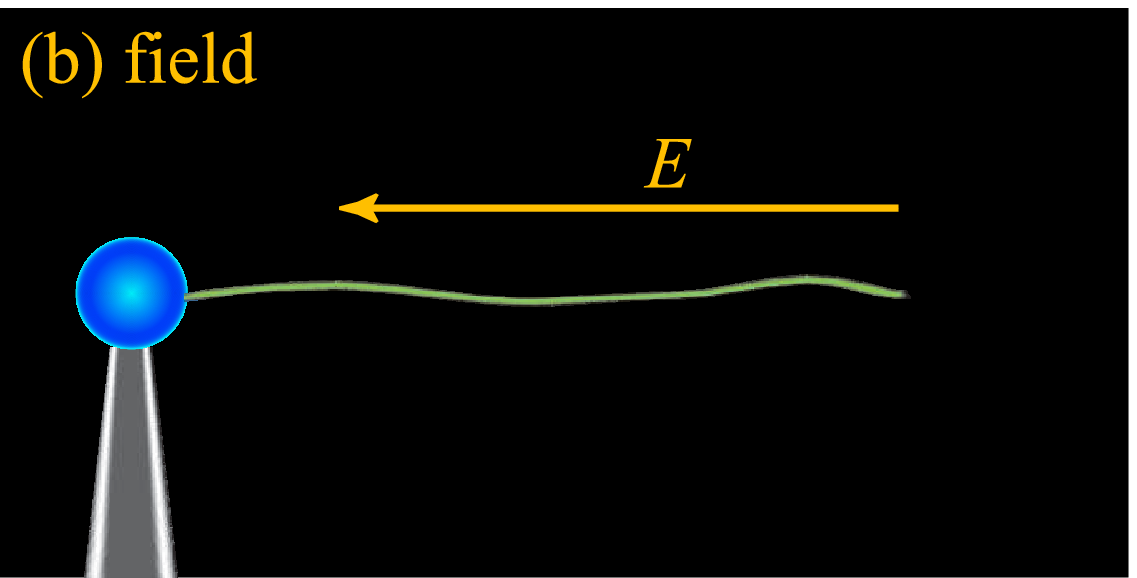}\\[.2cm]
  \includegraphics[width=.23\textwidth]{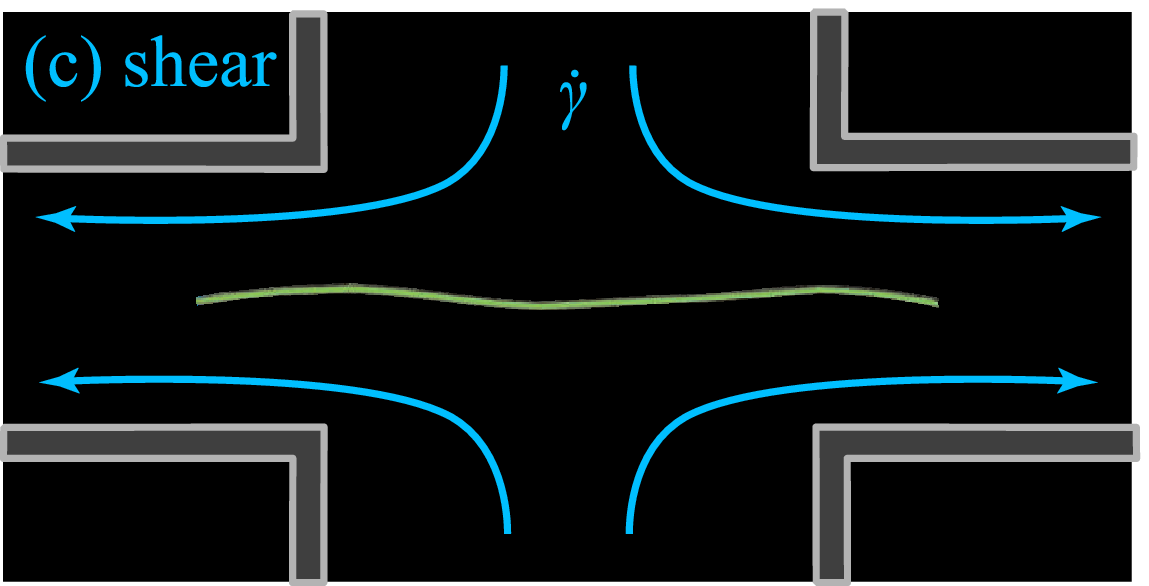}\hfill
  \includegraphics[width=.23\textwidth]{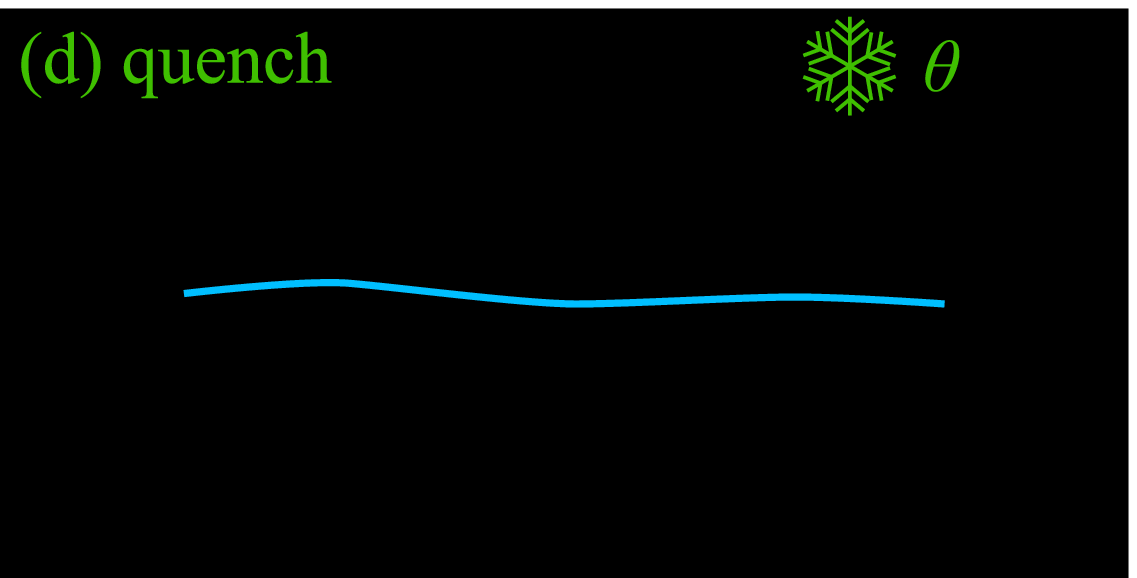}
  \caption{\label{fig:setups} (color online) In the ``force'' scenario
    (a), a stretching force $\fpre$ is suddenly removed at both
    ends. In the ``field'' setup (b), one end is held fixed and the
    electric field of strength $E$ (or homogeneous elongational flow
    of velocity $v$) is switched off, similar to the ``shear'' case
    (c) with a symmetric extensional shear flow of shear rate
    $\dot\gamma$.  In the ``quench'' experiment (d), the temperature
    is suddenly increased by a large factor $\theta$ from an initially
    small value.}
\end{figure}

In the following, we present results from computer simulations
combined with a thorough and exhaustive theoretical analysis to
explain how fundamental differences in the short-time relaxation
emerge from different experimental preparation methods but give way to
universal long-time relaxation.  Four idealized initial conditions
(see the cartoons in Fig.~\ref{fig:setups}) are shown to lead to
qualitatively distinct behavior despite superficial similarities.
\emph{``Force''} refers to mechanical stretching, i.e., a strong
external stretching force $\fpre$ that is suddenly removed on both
ends, for instance in a setup using $\lambda$-DNA, optical tweezers,
and restriction enzymes~\cite{vdbroek-noom-wuite:05}.  Secondly, the
term \emph{``field''} is used for experiments employing an electric
field~\cite{maier-seifert-raedler:02} of strength $E$ for stretching,
where one end is always kept fixed. Once switched off, such fields
give rise to relaxation dynamics similar to the one in setups using
homogeneous elongational flows~\cite{perkins-quake-smith-chu:94} of
velocity $v$.  Further, we denote by \emph{``shear''} the stretching
by planar extensional shear flows of shear rate $\dot\gamma$ in a
symmetric
geometry~\cite{schroeder-babcock-shaqfeh-chu:03,schroeder-shaqfeh-chu:04},
see Fig.~\ref{fig:setups}(c).  Finally, \emph{``quench''} refers to a
scenario where the temperature is suddenly increased by a large factor
$\theta$ from a small value $T/\theta$ near zero to its final value
$T$. This setup is more feasible for computer simulations, but the
equivalent sudden drop in persistence length $\lp$ might be
experimentally realizable by chemical reactions.

The paper is organized as follows. In the first section, we present
results from Brownian dynamics simulations for each of the four
different setups. The next section presents a qualitative discussion
of the underlying theoretical model resulting in scaling laws for
pertinent observables, which readily suggest intuitive explanations
for the qualitative differences between the scenarios and their
universal long-time asymptote. A detailed and somewhat technical
derivation of these asymptotic scaling laws is contained in the third
section, where we also analyze the effect of different boundary
conditions. In the fourth section, we present a quantitative
comparison between simulation results and theory.  At the end of the
paper, we discuss experimental implications including quantitative
estimates of control parameters in typical realizations, the onset of
the stretch--coil transition and the influence of hydrodynamic
interactions.

\section{Simulation results}

In the Brownian dynamics simulation, we employ the standard
free-draining bead-spring algorithm for wormlike chains, were
different environmental conditions during equilibration of the chains
correspond to the four scenarios introduced above. The equations of
motion for a chain of total length $L=N b$ with $N+1$ beads of size
$b$ and mobility $\mu$ are given by
\begin{equation}
  \pd_t\bvec r_i-\bvec v_i=-\mu\bnabla_i U + \bvec \eta_i(t),
\end{equation} where the potential $U=U_\text{s}+U_\text{b}+U_\text{f}$
contains a stretching part
\begin{equation}
  U_\text{s}=\frac{\kb T \gamma_\text{s}}{2 b}\sum_i
  (\abs{\bvec r_{i+1}-\bvec r_i}-b)^2,
\end{equation}
a bending part
\begin{equation}
  U_\text{b}= \frac{\kb T\lp}{b} \sum_i (1-\bvec t_{i+1}\cdot\bvec t_i),
\end{equation}
and an external potential $U_\mathrm{f}$. Here, $\lp$ is the
persistence length, $\gamma_\text{s}$ is the stretching elastic
constant, and $\bvec t_i=\frac{\bvec r_{i}-\bvec r_{i-1}}{\abs{\bvec
    r_{i}-\bvec r_{i-1}}}$ is a normalized tangent vector.  We use
Gaussian noise with strength $\ave{\bvec \eta_i(t)\bvec
  \eta_j(t')}=6\mu\kb T\delta_{ij}\delta(t-t')$. The time step is
$10^{-5}\tau_0$, where $\tau_0=b^2/(\kb T\mu)$ is the self-diffusion
time of the beads. The chains are equilibrated along the $x$-axis
symmetrically to the origin under the respective stretching
mechanism. In the ``force'' case, $U_\mathrm{f}=-\fpre (x_N-x_0)$ and
$\bvec v_i=0$, while $U_\mathrm{f}=0$ and $\bvec v_i = (v,0,0)^T$ for
``field'' setups.  For ``shear'', we take $\bvec
v_i=\dot\gamma\,(x_i,-y_i,-z_i)^T$, and in order to prevent the
polymer from diffusing out of the stagnation point, an additional
harmonic potential $U_\mathrm{f}=\frac{1}{2}\dot\gamma
\mu^{-1}x_\mathrm{COM}^2$ drives the center-of-mass coordinate
$x_\mathrm{COM}$ back to the origin (cf.\ the feedback control system
in Ref.~\cite{schroeder-babcock-shaqfeh-chu:03}).  In these scenarios,
we equilibrate for $10^4$ time steps, while initial conformations are
generated directly using the equilibrium tangent correlations in the
``quench'' case.  In all cases, $U_\mathrm{f}=0$ and $\bvec v_i=0$
upon release.  Ensemble averages were taken over 150 realizations

\begin{figure}
  \psfrag{X1}{$\tau/\tau_0$}
  \psfrag{X2}{$\tau/\tau_0$}
  \psfrag{Y1}{$\delta R_\parallel(t)/L$}
  \psfrag{Y2}{$\bar \fb(t)/(\kb T/b)$}
  \includegraphics[angle=270,width=.48\textwidth]{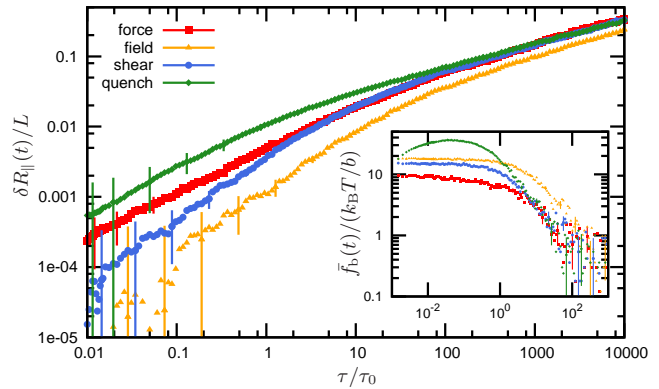}
  \caption{\label{fig:quantitative-results} (color online)
    Quantitative results: change in projected length $\delta
    R_\parallel(t)$ and mean bulk tension $\bar\fb(t)$ (inset) for
    computer simulations of a ``force'' (squares), ``field''
    (triangles), ``shear'' (circles) and ``quench'' (diamonds)
    scenario, respectively.  Simulation parameters were $L=200 b$,
    $\lp=40 b$, $\gamma_\text{s}=6000/b$, $\fpre=10\ \kb T/b$,
    $v=0.18\ b/\tau_0$, $\dot\gamma=0.0046/\tau_0$ and $\theta=35.7$,
    such that $R_\parallel(0)\approx 0.97 L$ in all cases.}
\end{figure}

To characterize the relaxation dynamics, we concentrate on two
observables.  One is the time dependent change $\delta
R_\parallel(t)=R_\parallel(0)-R_\parallel(t)$ in the ensemble average
of the filament's end-to-end distance $R_\parallel$, projected onto
the initial longitudinal axis. Note that with this definition, $\delta
R_\parallel$ is positive and increasing, while the actual end-to-end
distance shrinks during relaxation. The second observable is the mean
tension $\bar \fb(t)$ in the filament, proportional to the bulk stress
$\sigma(t)$ in a polymer solution.
Fig.~\ref{fig:quantitative-results} shows simulation results for
$\delta R_\parallel(t)$, measured from the projection on the initial
longitudinal axis, and for $\bar \fb(t)$ (proportional to the sum of
the spring displacements from their equilibrium position). Parameters
were chosen such that the initial extension is close to full
stretching ($R_\parallel(0)\approx 0.97L$ in all cases).  While the
universal scaling for longer times is evident (the apparent systematic
offset in the ``field'' case arises simply because there is only one
free end), substantial differences between the scenarios for shorter
times are clearly observable as well.

\section{Qualitative theoretical results}

From Fig.~\ref{fig:quantitative-results} it is obvious that the time
evolution of $\delta R_\parallel(t)$ and $\bar\fb(t)$ does not obey
simple power law scaling.  Simplifying approaches based on scaling
arguments~\cite{brochard:95,maier-seifert-raedler:02}, elastic
dumbbell models~\cite{ladoux-doyle:00,schroeder-shaqfeh-chu:04}, or
quasi-equilibrium
approximations~\cite{brochard-buguin-degennes:99,crut-etal:07} have
sometimes been used successfully for specific situations and parameter
ranges. In contrast, we employ a systematic
formalism~\cite{hallatschek-frey-kroy:05} based on the wormlike chain
model, which allows to generally account for the complex dynamics
resulting from different environmental perturbations.  Here, we first
present qualitative results for all four scenarios in order to
illustrate their differences, and discuss exact analytical and
numerical results in the next section.

\begin{figure}
    \includegraphics[width=.48\textwidth]{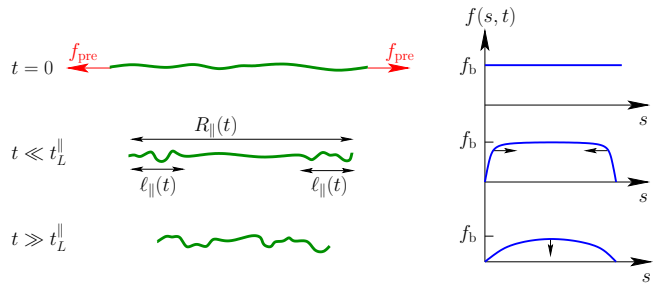}
    \caption{\label{fig:schema} (Color online) Schematic
      representation of conformational (left) and tension relaxation
      (right) in a ``force'' setup.  For $t=0$, the filament is
      equilibrated under the force $\fpre$.  As the ends are released,
      the contour coils up in two growing boundary layers of size
      $\lpar(t)$ where the tension relaxes.  At $t=\tlpar$, the
      dynamics crosses over from the propagation to the relaxation
      regime and the tension relaxes to zero.}
\end{figure}

In the wormlike chain model~\cite{saito-takahashi-yunoki:67},
semiflexible polymers are represented as inextensible smooth
spacecurves $\bvec r(s,t)$ of length $L$. Bending energy is
proportional to the squared local curvature $(\pd_s^2\bvec r)^2$, such
that in equilibrium, tangent orientations are correlated over the
persistence length $\lp=\kappa/\kb T$, where $\kappa$ is the bending
rigidity. The initially straight polymer is supposed to be
equilibrated at times $t<0$ and released at $t=0$.  After that, the
longitudinal contraction is driven energetically uphill via the
creation of contour undulations by entropic forces.  Considering the
conservation of contour length due to the (near) inextensibility of
the backbone bonds, these transverse wrinkles are conveniently
referred to in terms of their excess contour length, or \emph{stored
  length}, with an associated line density $\varrho$.  Mathematically,
the inextensibility is enforced by the \emph{backbone tension} $f$
which counteracts stretching, and the creation of stored length is
accompanied by the relaxation of tension.  The theory of
Refs.~\cite{hallatschek-frey-kroy:05,hallatschek-frey-kroy:07a}
relates the tension $f(s,t)$ to the stored length density
$\varrho(s,t)$, based on the \emph{weakly-bending} limit of small
contour deviations from a straight line.  In practice, this can easily
be realized by choosing the control parameters $\fpre$, $E$ or $v$, or
$\dot\gamma$ sufficiently strong (as in typical experiments, see
Table~\ref{tb:numbers} below), or the quenching factor $\theta$
sufficiently large. It also justifies the free-draining approximation,
where hydrodynamic effects are captured by anisotropic local friction
coefficients $\zeta_{\perp,\parallel}$ (per length) for
transverse/longitudinal friction, respectively~\cite{doi-edwards:86}.
However, ordinary perturbation theory is applicable only for late
times $t \gg t_\star\simeq \zeta_\parallel L^8/(\kb
T\lp^5)$~\cite{hallatschek-frey-kroy:05}, because to lowest order it
allows only a linear spatial dependence of $\varrho$ and $f$ and
neglects longitudinal friction forces~\cite{seifert-wintz-nelson:96}.
Further, except for quite stiff filaments with $\lp\gtrsim L$, the
time $t_\star$ is usually larger than the filament's longest
relaxation time $\tR\simeq\zeta_\parallel \lp L^2/\kb
T$~\cite{hallatschek-frey-kroy:07b}, which within our approximations
is given by the Rouse time of a polymer with Kuhn length $2\lp$.
Nevertheless, with an improved
formalism~\cite{hallatschek-frey-kroy:05,hallatschek-frey-kroy:07a}
including nontrivial spatial variations in $f$ and $\varrho$, the
conformational relaxation at times $t\ll \tR$ can be analyzed even for
quite flexible polymers.  This leads to the remarkable insight that
weakly-bending polymers constitute self-averaging systems: the small
stochastic fluctuations average out along the contour and the
coarse-grained tension dynamics follows from the deterministic
relation
\begin{equation}\label{eq:tension-dynamics}
  \pd_s^2 \bar f = -\zeta_\parallel \pd_t \ave{\bar\varrho},
\end{equation}
where the overbar denotes a (local) spatial average that produces
effectively an ensemble average (denoted by $\ave{\ .\
}$)~\cite{hallatschek-frey-kroy:07a}. Driven by tension gradients,
stored length propagates subdiffusively from the filament's ends into
the bulk---limited to boundary layers of size $\lpar(t)$ by
longitudinal solvent friction.  In more intuitive terms, the filament
starts to ``coil up'' first at the boundaries, and only later in the
bulk, see also Fig.~\ref{fig:schema}.

\begin{figure*}
  \begin{minipage}{\textwidth}
    \includegraphics[width=.32\textwidth]{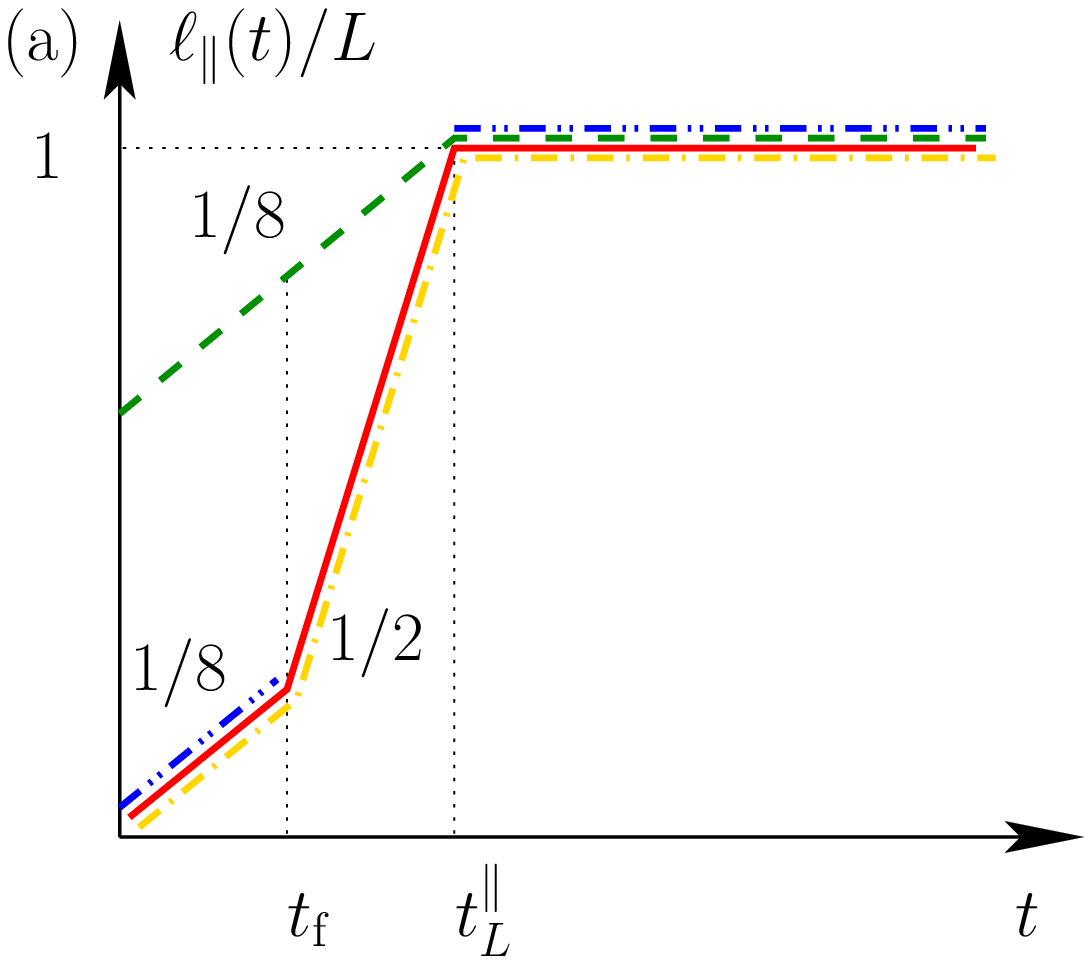}
    \includegraphics[width=.32\textwidth]{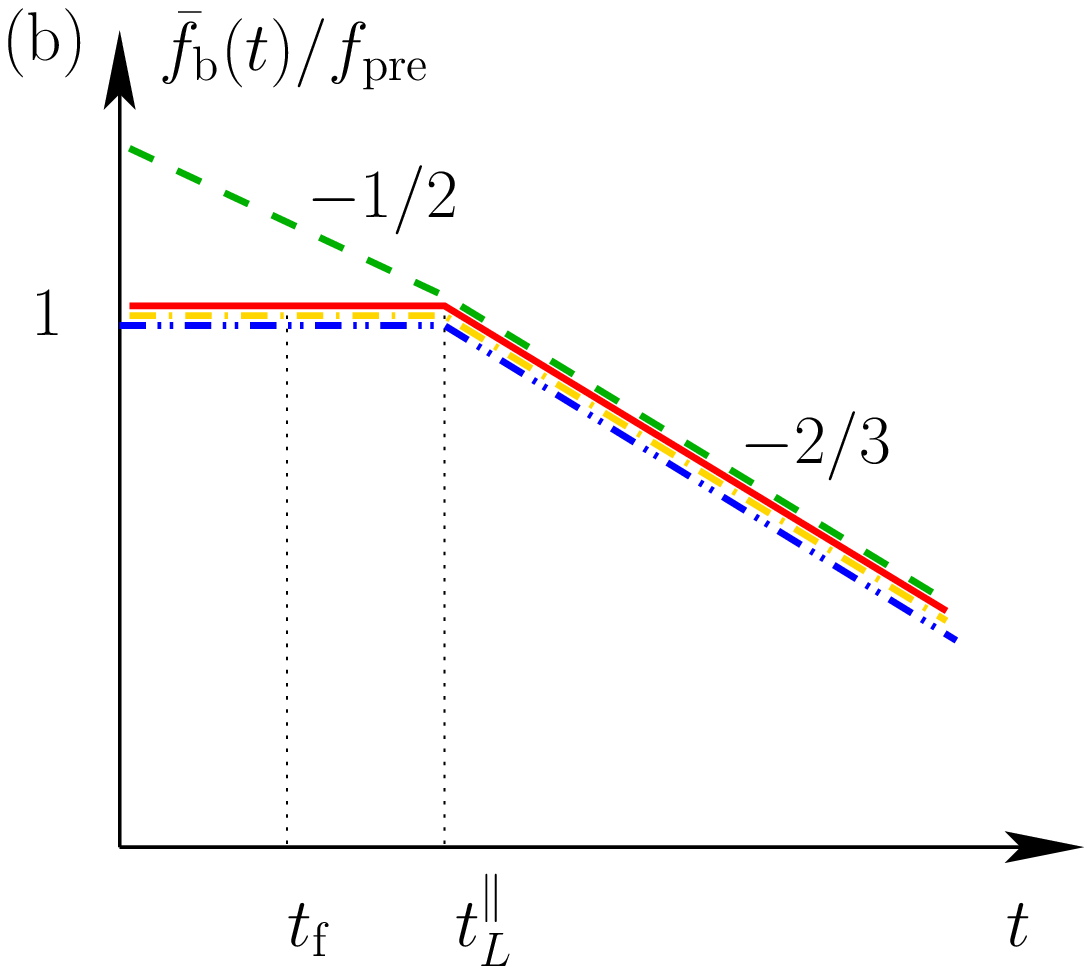}
    \includegraphics[width=.32\textwidth]{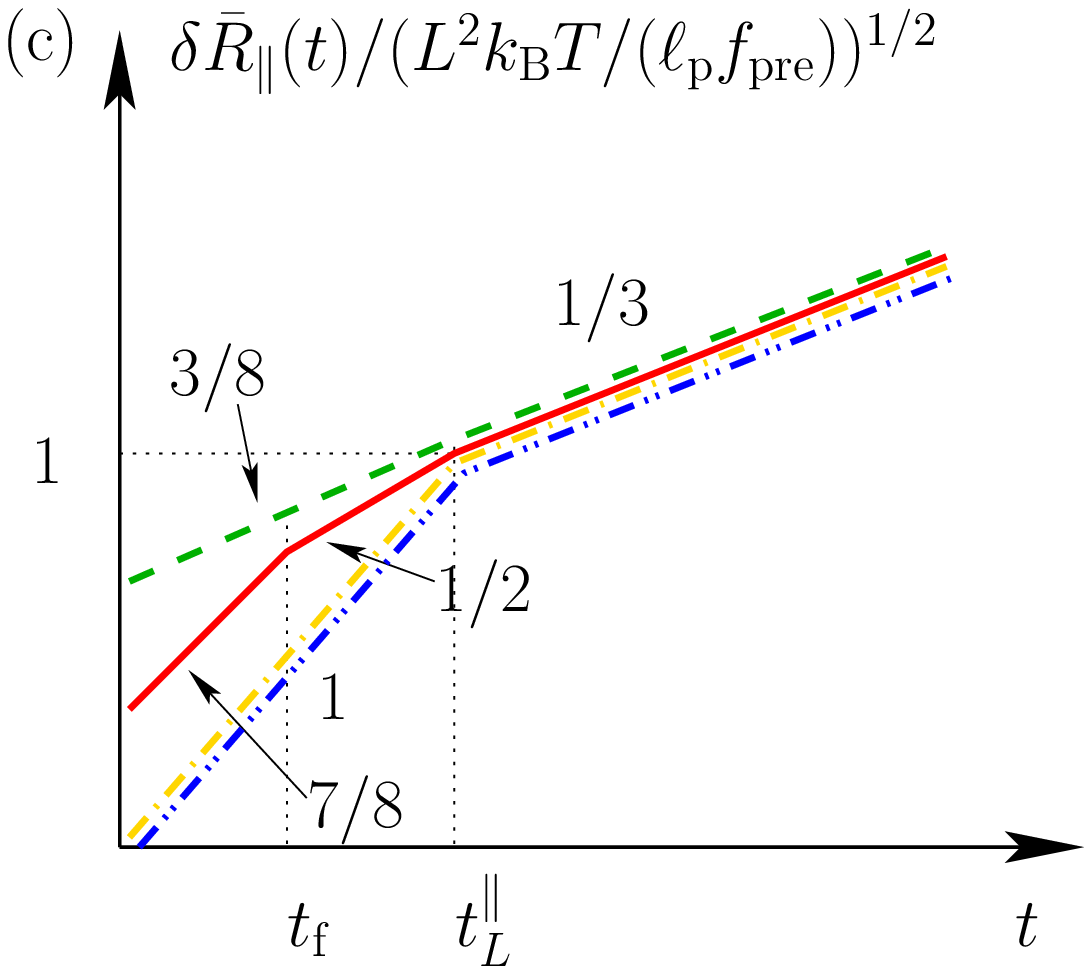}
  \end{minipage}
  \caption{\label{fig:qualitative-results} (color online) Qualitative
    results. Asymptotic scaling laws for the boundary layer size
    $\lp(t)$ (a), the mean bulk tension $\bar\fb(t)$ or the bulk
    stress (b), and for the change in projected length $\delta \bar
    R_\parallel(t)$ (c) in ``force'' (solid), ``field'' (dot-dashed),
    ``shear'' (dot-dot-dashed), or ``quench'' (dashed) setups,
    respectively.}
\end{figure*}

In general, $\ave{\bar\varrho}$ is a nonlinear functional of $\bar f$,
see Eq.~\eqref{eq:rho} below for a detailed expression. Exact
analytical results for the boundary layer size $\lpar(t)$, the bulk
tension $\bar \fb(t)$, and the change in projected length $\delta
R_\parallel(t)$ will be obtained as leading-order results of a
systematic asymptotic expansion of Eq.~\eqref{eq:tension-dynamics} in
the next section. However, the \emph{scaling} of the dominant part
$\delta\bar R_\parallel(t)$ of $\delta R_\parallel(t)$, which is
independent of boundary conditions and an effectively deterministic
quantity, can be found from a simple dimensional argument: the change
in end-to-end distance equals the amount of stored length
$\varrho\lpar$ that has been created in the boundary layer. On the
scaling level, Eq.~\eqref{eq:tension-dynamics} reads $\bar
\fb/\lpar^2\simeq \zeta_\parallel\varrho/t$, and we obtain
\begin{equation}\label{eq:observables}
  \delta \bar R_\parallel\simeq \frac{t\bar\fb}{\zeta_\parallel\lpar}.
\end{equation}
Note that the change $\delta R_G(t)$ in the gyration tensor's largest
eigenvalue, which is frequently identified with $\delta
R_\parallel$~\cite{nam-lee:07,dimitrakopoulos:04}, obeys a different
scaling law $\delta R_G\simeq t \bar\fb/(\zeta_\parallel L)$ for short
times $t\ll\tlpar$~\cite{hallatschek-frey-kroy:04}. Additional
subdominant contributions to $\delta R_\parallel(t)$ from end
fluctuations will be analyzed in
section~\ref{sec:quantitative-results}.

Fig.~\ref{fig:qualitative-results} summarizes the scaling results of
section~\ref{sec:quantitative-results} for $\lpar(t)$, $\bar \fb(t)$,
and $\bar\delta R_\parallel(t)$ in various intermediate asymptotic
regimes, which are separated by different crossover times that have
been matched by an appropriate choice of the respective control
parameters for better comparison.  Clearly, the time
$\tlpar=\zeta_\parallel L^2 [\kb T/(\lp\fpre^3)]^{1/2}$ is of key
importance since it separates scenario-specific and universal
relaxation. To understand the origin of the differences for times
$t\ll\tlpar$, we will consider the different scenarios separately
before we address the universal regime $t\gg\tlpar$.

\paragraph*{``Force'' setup.}
After the stretching force has been shut off, the polymer starts to
build up contour undulations driven by thermal noise.  These
transverse undulations appear first in growing boundary layers of size
$\lpar(t) \ll L$ near the ends: assuming an inextensible backbone, the
immediate creation of undulations in the bulk would require the ends
to be pulled inwards against longitudinal solvent friction with a
force exceeding the actual backbone tension.  This phenomenon of
\emph{tension propagation} ends after a time $\tlpar$, defined via
$\lpar(\tlpar)=L$, where the boundary layers extend over the whole
polymer length, see Fig.~\ref{fig:schema}. A more detailed
analysis~\cite{hallatschek-frey-kroy:05} shows that the
longitudinal relaxation depends on whether the internal tension
(initially equal to the stretching force $\fpre$) represents a
relevant perturbation to the transverse conformational dynamics. The
latter undergoes a dynamic crossover from a bending-dominated regime
with $\lpar\propto t^{1/8}$~\cite{everaers-juelicher-ajdari-maggs:99}
for the shortest times $t\ll\tf$ to a tension-driven regime with
$\lpar\propto t^{1/2}$~\cite{brochard-buguin-degennes:99} for longer
times $t\gg\tf$, where the dynamics becomes inherently nonlinear.
Here, $\tf=\zeta_\perp\kb T\lp/\fpre^2$ is a crossover time that obeys
$\tf\ll\tlpar$ for reasonably large prestress $\fpre$ (see
Table~\ref{tb:numbers} below).

\paragraph*{``Field'' and ``shear'' setup.} 
Similarly, one can define a force equivalent in ``field'' or ``shear''
experiments and corresponding expressions for $\tf$ and $\tlpar$: the
hydrodynamic equivalent of $\fpre$ is simply the total Stokes friction
$\zeta_\parallel v L$ in a homogeneous flow, and
$\zeta_\parallel\dot\gamma L^2$ is the longitudinal friction in an
extensional shear flow.  Flow conditions may straightforwardly be
recast into the equivalent language of external (e.g., electrical)
fields. However, complicated counterion
effects~\cite{long-viovy-ajdari:96a,stigter-bustamante:98,heuvel-graaff-lemay-dekker:07}
prevent the quantitative prediction of the equivalent electrophoretic
field strength $E$ for typical experimental realizations.  Although
field-type perturbations induce dynamic crossovers at $\tf$ similar to
the ``force'' case, the change $\delta R_\parallel(t)$ in projected
length increases always linearly with time.  This can be understood by
a simple change in perspective: the polymer's ends are pulled inwards
by an approximately constant bulk tension, i.e., with roughly constant
velocity.  This corresponds in the frame of reference of the ends to
an external \emph{flow} field. The resulting friction forces are
properly balanced and the initial polymer conformations are already
equilibrated under such a flow field in ``field'' and ``shear''
setups, in contrast to the ``force'' scenario.  Tension propagation is
therefore not a dominant effect in the former
(Eq.~\eqref{eq:observables} applies with $\lpar\equiv L$ because of
the large-scale spatial variation of the tension), and the constant
drag gives $\delta\bar R_\parallel\propto t$.

\paragraph*{``Quench'' setup.} 
Here, finally, there is no external force scale and therefore no
dynamic crossover ($\lpar\propto t^{1/8}$ for $t\ll \tlpar$), although
the parameter combination $\kb T \lp^3\theta^4/L^4$ plays the role of
$\fpre$ in the crossover time $\tlpar$.  The tension in a quenched
filament is produced solely by the suddenly increased thermal noise
from the environment (if the external temperature increases), or by
the suddenly higher ``sensitivity'' to this noise (if the quenching is
achieved by a sudden drop in bending rigidity).  Hence, its magnitude
depends on the ``mismatch'' between the current conformation and an
equilibrium conformation corresponding to the current environment.
Therefore, the quenched filament can relax tension \emph{even in the
  bulk}, by reshuffling stored length between long and short
wavelength modes in a way similar to the mechanical stress relaxation
in buckled rods~\cite{hallatschek-frey-kroy:04}, while the bulk
tension stays constant for $t\ll\tlpar$ in the other setups.

\paragraph*{Universal regime.} 
At long times $t\gg\tlpar$, when the tension has propagated through
the filament, the dynamics enters the universal regime of homogeneous
tension relaxation~\cite{hallatschek-frey-kroy:05}.  Contrary to
previous assumptions~\cite{bohbot_raviv-etal:04}, longitudinal
friction may not generally be neglected, but dominates the dynamics in
this regime.  The tension has a nontrivial spatial dependence, but it
can for asymptotically large forces be treated as quasi-statically
equilibrated~\cite{brochard-buguin-degennes:99,hallatschek-frey-kroy:05}.
The characteristic universality of the long-time relaxation is then
simply a consequence of the right hand side of
Eq.~\eqref{eq:tension-dynamics} being independent of initial
conditions: $\bar\varrho\approx [\kb T/(4 \lp \bar f)]^{1/2}$, where
we have used the (static) force-extension relation for wormlike
chains~\cite{marko-siggia:95}. This asymptote readily implies by
Eq.~\eqref{eq:tension-dynamics} the scaling $\bar f\propto t^{-2/3}$
and by Eq.~\eqref{eq:observables} the characteristic $t^{1/3}$-growth
of $\delta \bar R_\parallel(t)$, which has indeed been observed in
experiments~\cite{schroeder-shaqfeh-chu:04}.  As an aside, we note
that $t_\star < \tR$ for stiff polymers with $L\lesssim\lp$; the
adjoining regime of algebraic relaxation for times $t_\star\ll t$
shows a $t^{1/4}$-scaling in $\delta\bar
R_\parallel(t)$~\cite{granek:97,hallatschek-frey-kroy:05}.

Let us finally comment on the joint limiting scenario: the
\emph{exactly} straight initial conformation (as in
Ref.~\cite{dimitrakopoulos:04}). Not only it is quite artificial from
a theoretical and experimental perspective, it also appears to be
ambiguous, since we could let $\fpre\to\infty$ in one of the scenarios
involving external forces as well as $\theta\to\infty$ for the
``quench'' case.  Although $\tlpar\to 0$ in both cases, so that only
the universal regime survives and the ambiguity is limited to $t=0$,
it gives rise to observable effects as soon as one takes into account
some microstructure corrections important for real experimental
systems and simulation models~\cite{obermayer:unpub}.

After this qualitative discussion of the relaxation dynamics, we will
now present a systematic derivation and analysis of
Eq.~\eqref{eq:tension-dynamics}, resulting in exact growth laws in
various intermediate asymptotic regimes for the observables introduced
above. While Ref.~\cite{hallatschek-frey-kroy:07b} only covered the
``force'' case, we now obtain results for the other scenarios as well,
and include a quantitative analysis of different boundary conditions.

\section{\label{sec:quantitative-results}Quantitative theoretical
  results}

Starting point for our calculations is the wormlike-chain Hamiltonian
\begin{equation}\label{eq:wlc-hamiltonian}
  \mathcal{H}=\frac{1}{2} \int_0^L\!\td s\left[\kappa \bvec r''^2 + f \bvec r'^2\right],
\end{equation}
where the backbone tension $f(s,t)$, a Lagrange multiplier
function~\cite{goldstein-langer:95}, takes care of the local
inextensibility constraint $\bvec r'(s)^2=1$. The equations of motion
for the contour result from balancing elastic forces $-\pd
\mathcal{H}/\pd \bvec r$ with stochastic noise $\bvec \xi$ and
anisotropic viscous local friction forces $\bvec\zeta[\pd_t\bvec
r-\bvec u]$ with friction matrix $\bvec\zeta=[\zeta_\perp \bvec
r'\bvec r'+\zeta_\parallel (1-\bvec r'\bvec r')]$ and a velocity field
$\bvec u$ of the solvent. The friction coefficients (per length) are
$\zeta_\parallel=\hat\zeta\zeta_\perp$ with $\hat\zeta\approx 1/2$ and
$\zeta_\perp\approx4\pi\eta/\ln(L/a)$~\cite{doi-edwards:86}, where $a$
is the backbone thickness. In all of this section, we set
$\zeta_\perp=\kappa\equiv 1$ for simplicity. This makes time a
length$^{4}$ and the tension a length$^{-2}$. Note that $\kappa=\kb
T\lp$ is kept constant in the ``quench'' scenario. Our approach
exploits the weakly-bending limit.  Parameterizing the contour $\bvec
r= (\rperp, s-\rpar)^T$ in terms of small transverse and longitudinal
displacements from the straight ground state, this means that
$\rperp'^2=\mathcal{O}(\eps)\ll 1$, with $\eps=\fpre^{-1/2}/\lp$ for
``force'' setups (and $\fpre$ replaced by its equivalents in ``field''
or ``shear'' scenarios) and $\eps=L/(\theta\lp)$ for ``quench''
setups, respectively. Up to order $\eps$, the equations of motion for
the contour in absence of external forces and for $\bvec u=0$ read:
\begin{subequations}\label{eq:eom}
  \begin{align}
    \label{eq:eom-T}
    \pd_t\rperp &= -\rperp'''' + (f\rperp')' + \bvec \xi_\perp \\
    \label{eq:eom-L}
    \hat\zeta\pd_t \rpar + (1-\hat\zeta)\rperp'\pd_t\rperp &=
    -\rpar'''' - f' + (f\rpar')' + \xi_\parallel
  \end{align}
\end{subequations}
Because in the weakly-bending limit the transverse contour
fluctuations are correlated on much shorter length scales than the
longitudinal (= tension) dynamics, we can formally introduce ``fast''
and ``slow'' arclength coordinates for the small-scale transverse and
large-scale longitudinal dynamics,
respectively~\cite{hallatschek-frey-kroy:07a}.  Taking a local (with
respect to $\lpar$) spatial average over the small-scale fluctuations
(denoted by an overbar) leads to closed equations:
\begin{subequations}
  \label{eq:eom-mspt}
  \begin{align}
    \label{eq:eom-mspt-T}
    \pd_t \rperp &= -\rperp'''' + \bar f\rperp'' + \bvec \xi_\perp \\
    \label{eq:eom-mspt-L}
    \pd_s^2 \bar f &= -\hat\zeta\pd_t\ave{\bar{\varrho}}.
  \end{align}
\end{subequations}
The longitudinal part Eq.~\eqref{eq:eom-mspt-L}, where
$\varrho=\tfrac{1}{2}\rperp'^2$ is the stored length density, follows
from the self-averaging property of weakly-bending polymers: the
spatial coarse-graining effectively generates an ensemble average. The
transverse part Eq.~\eqref{eq:eom-mspt-T} contains a locally constant
tension $\bar f$ (its slow arclength dependence obtained through
Eq.~\eqref{eq:eom-mspt-L} is adiabatically inherited), and can be
solved in terms of appropriate eigenmodes $w_q(s)$ with eigenvalue
$-q^2 (q^2+f(t))$ via the response function
\begin{equation}\label{eq:chi-perp}
  \chi_\perp(q;t,t')=\e^{-2 q^2 [ q^2 (t-t') + \int_{t'}^t\!\td \tau \bar f(\tau)]},
\end{equation} 
Using the noise correlation $\ave{\bvec \xi_\perp(k,t)\bvec
  \xi_\perp(q,t')}=4\lp^{-1} \delta_{k,q}\delta(t-t')$, we evaluate
the expectation value $\ave{\tfrac{1}{2}\rperp'^2}$. The different
preparation mechanisms discussed in the main text constrain the
polymer only for $t<0$.  Including the initial conditions $\bar
f(s,t<0)=\bar f_0(s)$ and $\lp(t > 0)\equiv \lp= \lp(t<0)/\theta$
gives for the stored length density
\begin{multline}\label{eq:rho}
  \ave{\varrho} = \frac{1}{\lp}\sum_q \Bigg[\frac{\chi_\perp^2(q;t,0)}
  {q^2 \theta(q^2 + \bar f_0(s))} \\
  + 2 \int_0^t\!\td t'\,\chi_\perp^2(q;t,t')\Bigg] w_q'^2(s)
\end{multline}
As only the spatially averaged stored length density $\ave{\bar
  \varrho}$ enters Eq.~\eqref{eq:eom-mspt-L}, we decompose $w_q'^2(s)$
into a spatially constant and a fluctuating part $c_q(s)$ (the latter
will average out upon coarse-graining):
\begin{equation}\label{eq:mode-decomposition}
  w_q'^2(s) = \frac{q^2}{L}\left[ 1 + c_q(s)\right].
\end{equation}
Taking the continuum limit $L\to\infty$ and integrating
Eq.~\eqref{eq:eom-mspt-L} over time, we find that the integrated
tension $\bar F(s,t)=\int_0^t\!\td t' \bar f(s,t')$ obeys the partial
integro-differential equation
\begin{multline}\label{eq:pide}
  \pd_s^2 \bar F(s,t) = \hat\zeta \int_0^\infty\!\frac{\td
    q}{\pi\lp}\Bigg[ \frac{1-\chi_\perp^2(q;t,0)}{\theta(q^2+\bar
    f_0(s))}\\-2 q^2 \int_0^t\!\td t'\chi_\perp^2(q;t,t')\Bigg].
\end{multline}
From solutions to this equation in different intermediate asymptotic
regimes presented in the next subsection, we will then infer growth
laws for the two observables.

\subsection{\label{sec:asymptotic-tension}Asymptotic results for the
  tension}

\subsubsection{\label{sec:force}``Force'' setup}

This scenario with $\theta=1$ and the initial and boundary conditions
\begin{equation}\label{eq:bc-force}
  \begin{split}
    \bar f(s,t<0)=\fpre \quad \text{ and } \\
    \bar f(0,t >0)=0,\quad \bar f(L,t>0)=0,
  \end{split}
\end{equation}
is identical to the ``release''-scenario which was thoroughly analysed
in Ref.~\cite{hallatschek-frey-kroy:07b}. We will briefly sketch this
analysis in order to motivate its application to the other
setups. From the response function Eq.~\eqref{eq:chi-perp}, we get the
asymptotic scaling for the wave number $\qm$ of the mode that relaxes
at time $t$:
\begin{equation}\label{eq:qm}
  \qm \simeq \begin{cases}
    t^{-1/4},&\quad\text{if }\bar F^2/t\ll 1\quad\text{(``linear'')}\\
    \bar F^{-1/2},&\quad\text{if }\bar F^2/t\gg 1\quad\text{(``nonlinear'')}
  \end{cases}\end{equation}
Examining Eq.~\eqref{eq:eom-mspt-T}, one infers that in the first case
the tension contribution $\qm^2 \bar F$ is small compared to the
bending contribution $\qm^4 t$ and can be treated as perturbation on
the linear level. Since the magnitude of the tension is determined by
the prestretching force, $\bar F\simeq \fpre t$, this asymptote,
called ``linear regime'', can also be formulated as $t\ll\tf$ with
$\tf=\fpre^{-2}$.  In the second case $t\gg\tf$, the bending
contributions are subdominant which leads to different ``nonlinear
regimes''.

\paragraph*{Linear propagation ($t\ll \tf$).} We perform an
expansion~\cite{obermayer-hallatschek-frey-kroy:07} of the right hand
side of Eq.~\eqref{eq:pide} with respect to the integrated tension
$\bar F$ and to the force $\fpre$:
\begin{multline}\label{eq:pide-linearized}
  \pd_s^2 \bar F(s,t) \approx \hat\zeta\int_0^\infty\!\frac{\td
    q}{\pi\lp}\Bigg[
  -\frac{\fpre}{q^4}\left(1-\e^{-2 q^4 t}\right) \\
  + 2 \bar F(s,t) - 4 q^4 \int_0^t\!\td t'\,\bar F(s,t')\e^{-2 q^4
    (t-t')}\Bigg].
\end{multline}
Using the Laplace transform $\tilde F(s,z)=\mathcal{L}\{\bar
F(s,t)\}$, this reads
\begin{equation}\label{eq:pide-linearized-laplace-0}
  \pd_s^2\tilde F(s,z)=\hat\zeta\int_0^\infty\!\frac{\td q}{\pi\lp}\left[
    -\frac{2 \fpre}{z(z+2q^4)} + \tilde F(s,z) \frac{2 z}{z+2 q^4}\right],
\end{equation}
which, after performing the $q$-integral, reduces to:
\begin{equation}\label{eq:pide-linearized-laplace}
  \lambda^2\,\pd_s^2\tilde F = \tilde F-\frac{\fpre}{z^2}.
\end{equation}
Here, $\lambda(z)=2^{3/8}(\lp/\hat\zeta)^{1/2} z^{-1/8}$ is a dynamic
length scale denoting the size of spatial variations in $\tilde
F(s,z)$. If $L\gg\lambda$, the solution to
Eq.~\eqref{eq:pide-linearized-laplace} varies only close to the
boundaries, as it is characteristic for the propagation regime. Near
$s=0$ (and correspondingly near $s=L$), it simplifies to
\begin{equation}\label{eq:laplace-solution-force-propagation}
  \tilde F(s,z) \approx\frac{\fpre}{z^2}\left[1- \e^{-s/\lambda}\right],
\end{equation}
which can be backtransformed~\cite{hallatschek-frey-kroy:07b} to
\begin{equation}\label{eq:solution-linear-propagation-force}
  \bar F(s,t) = \fpre\,t\left[1-\phi(s/\lpar(t))\right],
\end{equation}
where $\phi(\xi)\approx \exp[-2^{-3/8}\xi/\EGamma{15/8}]$ is a scaling
function that depends only on the ratio $\xi=s/\lpar(t)$.  The length
scale $\lambda$ is directly related to the boundary layer size
$\lpar(t)=(\lp/\hat\zeta)^{1/2}t^{1/8}$~\cite{everaers-juelicher-ajdari-maggs:99,hallatschek-frey-kroy:05,hallatschek-frey-kroy:07b},
and the requirement $L\gg\lambda$ translates into $t\ll\tlpar$.

\paragraph*{Nonlinear propagation ($\tf\ll t\ll\tlpar$).}
In the nonlinear regime, the $\qm^4 t$ bending contributions are small
compared to the tension terms $\qm^2 \bar F$ if $\bar F^2/t\gg
1$. This results in $\chi_\perp(q;t,t')$ being finite only near
$t'\approx t$, see Eq.~\eqref{eq:chi-perp}. We can therefore linearize
$\bar F(s,t)-\bar F(s,t')\approx [\pd_t \bar F(s,t)](t-t')$ in the
exponent. The $t'$-integral in the second term of Eq.~\eqref{eq:pide}
is readily performed~\cite{hallatschek-frey-kroy:07b}:
\begin{align}
  \pd_s^2 \bar F &\approx \hat\zeta\int_0^\infty\!\frac{\td
    q}{\pi\lp}\left[\frac{1-\chi_\perp^2(q;t,0)}
    {q^2+\fpre}-\frac{1-\chi_\perp^2(q;t,0)}{q^2 + \pd_t \bar
      F}\right]
  \nonumber \\
  &\approx \hat\zeta\int_0^\infty\!\frac{\td
    q}{\pi\lp}\left[\frac{1}{q^2+\fpre}-
    \frac{1}{q^2 + \pd_t\bar  F}\right] \nonumber \\
  \label{eq:pide-qs}
  &= \frac{\hat\zeta}{2\lp}\left[\fpre^{-1/2}-(\pd_t \bar
    F)^{-1/2}\right].
\end{align}
In the second line, we let $\chi_\perp\to 0$ because $\bar F^2/t\gg
1$.  This indicates the underlying ``quasi-static'' approximation: the
relevant modes have already decayed and the tension is
quasi-statically equilibrated.  Taking a time derivative
gives~\cite{brochard-buguin-degennes:99,hallatschek-frey-kroy:07b}
\begin{equation}\label{eq:pde-bbg}
  \pd_s^2 \bar f = \frac{\hat\zeta\pd_t \bar f}{4\lp \bar f^{3/2}}.
\end{equation}
Inspired by the result
Eq.~\eqref{eq:solution-linear-propagation-force}, we expect a scaling
form $\bar f(s,t)=\fpre \varphi(\xi)$ with $\xi=s/\lpar(t)$ for the
tension.  Inserting it into Eq.~\eqref{eq:pde-bbg} gives
$\lpar(t)=(\lp/\hat\zeta)^{1/2}\fpre^{3/4}t^{1/2}$~\cite{brochard-buguin-degennes:99,hallatschek-frey-kroy:05}
and
\begin{equation}\label{eq:ode-bbg}
  \pd_\xi^2\varphi = -\frac{1}{8}\xi\varphi^{-3/2}\pd_\xi\varphi,
\end{equation}
with the boundary conditions $\varphi(0)=0$ and
$\pd_\xi\varphi(\xi\to\infty)=0$, i.e., we neglect the presence of the
second end, where the situation is correspondingly, and assume just a
flat profile in the bulk. Numerical solutions to this equation have
been shown in
Refs.~\cite{brochard-buguin-degennes:99,hallatschek-frey-kroy:07b} and
give $\varphi(\xi\to\infty)=1$ as expected and $\varphi'(0)\approx
0.62$. The propagation regime ends at $\tlpar=L^2\fpre^{-3/2}/\lp$.

\paragraph*{Homogeneous relaxation ($\tlpar\ll t\ll \tR$).}
After the tension has propagated through the filament, it is no longer
constant but expected to decay.  But as long as $\bar F^2/t\gg 1$
still holds, we can use Eq.~\eqref{eq:pde-bbg}.  Hence, we try the
separation ansatz $\bar f(s,t)=g(t)h(\xi)$ with $\xi=s/L$, which
gives~\cite{hallatschek-frey-kroy:07b}
\begin{equation}\label{eq:solution-homogeneous-relaxation-force-g}
  g(t)=\left(\frac{\hat\zeta L^2}{\lp t}\right)^{2/3},
\end{equation}
and 
\begin{equation}\label{eq:solution-homogeneous-relaxation-force-h}
  h''=-\tfrac{1}{6} h^{-1/2}\quad \text{with }h(0)=h(1)=0.
\end{equation}
The almost parabolic profile $h(\xi)$ is characterized
by~\cite{hallatschek-frey-kroy:07b}
\begin{equation}\label{eq:solution-homogeneous-relaxation-force-h-characteristics}
  h'(0)=12^{-1/3},\quad h(1/2)=\left(\tfrac{3}{128}\right)^{2/3}.
\end{equation}
Using Eq.~\eqref{eq:solution-homogeneous-relaxation-force-g}, we find
that the condition $\bar F^2/t\gg 1$ is violated for $t\gtrsim
t_\star= L^8/\lp^4$, which is already larger than $\tR$ if
$\lp\lesssim L$.  Hence, this regime lasts until the weakly-bending
approximation breaks down near the ends due to the onset of the
stretch--coil transition.

\subsubsection{\label{sec:field}``Field'' setup}

For hydrodynamic and/or electrophoretic forces, we find from the
longitudinal equation of motion Eq.~\eqref{eq:eom-L} a corresponding
non-uniform initial tension profile $\bar f(s,t<0)=g (L-s)$ with
$g=\hat\zeta v$ for flows or $g\propto E$ for an electric field, where
the generally unknown prefactor is some combination of electrophoretic
and hydrodynamic mobility.  This linearly decreasing prestress would
in principle lead to an additional term $\bar f' \rperp'$ in
Eq.~\eqref{eq:eom-mspt-T}, and the corresponding eigenfunctions would
be very complicated. However, because large scale tension
variations are irrelevant for the short wavelength transverse
dynamics, we can ignore this term by consistently exploiting the scale
separation which allowed the derivation of Eq.~\eqref{eq:eom-mspt},
and use Eq.~\eqref{eq:pide} with the initial linear profile $\bar
f_0(s)=g(L-s)$. The polymer is supposed to be grafted at $s=0$ and to
have a free end at $s=L$, i.e., the boundary conditions are
\begin{equation}\label{eq:bc-field}
  \bar f'(0,t>0)=0 \quad\text{and}\quad \bar f(L,t>0)=0.
\end{equation}
Identifying the force equivalent $\fpre^*=g L$, we expect a linear
regime for $t\ll \tf$ with $\tf=\fpre^{*-2}$ and a nonlinear regime
for $t\gg \tf$, governed by the respective asymptotic differential
equations from the ``force'' case.

\paragraph*{Linear propagation ($t \ll \tf$).}
Linearizing Eq.~\eqref{eq:pide} in $\bar F$ and $\bar f_0(s)$ and
performing a Laplace transform as in
Eqs.~(\ref{eq:pide-linearized}--\ref{eq:pide-linearized-laplace}), we
arrive at the solution
\begin{equation}\label{eq:solution-linear-propagation-field}
  \bar F(s,t) = g t (L-s) - g \lpar(t) t \phi(s/\lpar(t)),
\end{equation}
with $\lpar(t)= (\lp/\hat\zeta)^{1/2} t^{1/8}$ and $\phi(\xi)\approx
\tfrac{2^{3/8}\exp[-\frac{\EGamma{17/8}}{2^{3/8}}\xi]}{\EGamma{17/8}}$.
We find a boundary layer \emph{at the fixed} end where the tension
relaxes from its initial value $g L$ only by the small amount
$g\lpar$. Near the free end, at $s=L$, we have $F(L,t)=0$ and
$F'(L,t)=-g t$ \emph{without} any algebraic correction terms. Hence,
because the tension at the free end is already very small and the
contour does not further coil up, there are no boundary layer effects
which would give relevant deviations from the linear drift towards the
grafted end, in contrast to what has been found in
Ref.~\cite{maier-seifert-raedler:02}.

\paragraph*{Nonlinear propagation ($\tf\ll t \ll\tlpar$).} The
assumption $F^2/t\gg 1$ leads again to Eq.~\eqref{eq:pde-bbg} except
for very small regions near the free end where $\bar f_0(s)$ in the
denominator of the first term of Eq.~\eqref{eq:pide} is almost zero.
Corresponding to the linear case
Eq.~\eqref{eq:solution-linear-propagation-field}, we assume that the
tension deviates only near the fixed end from its initial value
$\bar f_0(s)$.  Hence, we insert $\bar f(s,t)=\bar f_0(s)-g
\lpar(t)\varphi(\xi)$ with $\xi=s/\lpar(t)$ and
$\lpar(t)=(\lp/\hat\zeta)^{1/2} (g L)^{3/4} t^{1/2}$ into
Eq.~\eqref{eq:pde-bbg}, and expand about $\bar f_0(s\ll L)$:
\begin{equation}\label{eq:ode-bbg-field}
  \pd_\xi^2\varphi(\xi) \approx \frac{1}{8}\left[\varphi(\xi)-\xi\pd_\xi\varphi(\xi)\right]
\end{equation}
with $\varphi'(0)=-1$ and $\varphi (\xi\to\infty)=0$.  The solution
can be given in terms of the complementary error function:
\begin{equation}\label{eq:solution-nonlinear-propagation-field}
  \varphi(\xi)=\frac{4}{\sqrt{\pi}}\e^{-\xi^2/16} - \xi\erfc(\xi/4).
\end{equation}
The propagation regime ends at $\tlpar=L^2/(\lp (g L)^{3/2})$.
\paragraph*{Homogeneous relaxation ($\tlpar\ll t \ll \tR$).} The
separation ansatz $\bar f(s,t)=g(t)h(\xi)$ with $\xi=s/L$ in
Eq.~\eqref{eq:pde-bbg} gives $g(t)=(\hat\zeta L^2/(\lp t))^{2/3}$ as
in Eq.~\eqref{eq:solution-homogeneous-relaxation-force-g} and the
spatial function $h$ solves
\begin{equation}\label{eq:solution-homogeneous-relaxation-field-h}
  h'' = -\tfrac{1}{6} h^{-1/2}\quad\text{with }h'(0)=0,\quad h(1)=0.
\end{equation}
We find the following characteristics
\begin{equation}\label{eq:solution-homogeneous-relaxation-field-h-characteristics}
  h(0)=(\tfrac{3}{32})^{2/3},\quad h'(1)= 6^{-1/3}.
\end{equation}
The condition $\bar F^2/t\gg 1$ holds until $t=\tR$.

\subsubsection{\label{sec:shear} ``Shear'' setup}

In this scenario, the equations of motion $\bvec\zeta[\pd_t\bvec
r-\bvec u]=-\delta\mathcal{H}/\delta\bvec r + \bvec \xi$ are modified
in the presence of an extensional shear flow field $\bvec
u=\dot\gamma(-\rperp,s-\rpar-L/2)^T$, where $\dot\gamma$ is the shear
rate. To lowest order in $\eps$, and in the stationary state, we
obtain from Eq.~\eqref{eq:eom-L}
\begin{equation}\label{eq:eom-L-shear}
  -\hat\zeta\dot\gamma \left(s-\tfrac{L}{2}\right)
  = \bar f'.
\end{equation}
As before, we treat this non-uniform tension profile only as
large-scale variation and use Eq.~\eqref{eq:pide} with the initial and
boundary conditions
\begin{equation}\label{eq:bc-shear}
  \begin{split}
    \bar f(s,t<0)=\tfrac{1}{2}\hat\zeta \dot\gamma s\,(L-s) \text{ and} \\
    \bar f(0,t>0)=\bar f(L,t>0)=0.
  \end{split}
\end{equation}
As in the ``field'' case, the time $\tf=\fpre^{*-2}$ with the force
equivalent $\fpre^*=\hat\zeta\dot\gamma L^2$ denotes the
linear--nonlinear crossover.

\paragraph*{Linear propagation ($t\ll\tf$).}

Here the solution to Eq.~\eqref{eq:pide-linearized-laplace} reads
\begin{equation}\label{eq:solution-linear-propagation-shear}
  \bar F(s,t)= \tfrac{1}{2}\hat\zeta\dot\gamma t s(L-s) - \frac{2^{3/4}\hat\zeta\dot\gamma\lpar^2 t}{\EGamma{9/4}} \left[1-\phi(s/\lpar(t))\right],
\end{equation}
with $\lpar=(\lp/\hat\zeta)^{1/2}t^{1/8}$ as before and
$\phi(\xi)\approx \exp[-2^{-3/8}\EGamma{9/4} \xi/\EGamma{17/8}]$. We
find two small boundary layers at the ends where the tension is
slightly smaller than initially.

\paragraph*{Nonlinear regime ($\tf\ll t\ll \tlpar$).}

The nonlinear regime $\tf\ll t\ll\tlpar$ for the ``shear'' setup is
quite peculiar: if we try (similar to the ``force'' and ``field''
case) a scaling ansatz $\bar f(s,t)=\frac{1}{2}\hat\zeta\dot\gamma[ s
(L-s) + L \lpar(t) \varphi(s/\lpar(t))]$ or similarly, we get
$\lpar(t)\sim t^2$.  This unusual result could be explained by the
fact that the ``prestress'' $\bar
f(s,t<0)\approx\frac{1}{2}\hat\zeta\dot\gamma s L$, which is
responsible for the scaling of $\lpar$ in this
regime~\cite{obermayer-hallatschek-frey-kroy:07}, grows linearly with
the distance from the ends. However, we dot not get any physically
meaningful differential equation for $\varphi$ under the boundary
conditions~Eq.~\eqref{eq:bc-shear}. We conclude that there is no
propagation and \emph{no observable boundary layers}. Looking for a
solution spanning the whole arclength interval from $0$ to $L$
instead, we insert into Eq.~\eqref{eq:pde-bbg} an expansion of the
form
\begin{equation}
  \bar f(s,t)=\tfrac{1}{2}\hat\zeta\dot\gamma L^2 \left[\varphi_0(\xi) + \frac{t}{\tlpar} \varphi_1(\xi) + \mathcal{O}\left((t/\tlpar)^2\right)\right]
\end{equation}
with $\xi=s/L$ and $\tlpar=\hat\zeta L^2/[\lp (\hat\zeta\dot\gamma
L^2)^{3/2}]$.  Solving the resulting differential equations for
successive powers of $(t/\tlpar)$ gives the leading order terms
\begin{equation}
  \varphi_0(\xi)=\xi(1-\xi),\quad \varphi_1(\xi) = -[2\xi(1-\xi)]^{3/2}.
\end{equation}
In contrast to the propagation forms $\bar f(s,t)\sim
\varphi(s/\lpar(t))$ of the other scenarios, we now get self-similar
and spatially invariant tension profiles. This can probably attributed
to this specific initial condition which allows for self-similar
relaxation.  Further, to linear order in $(t/\tlpar)$ we do not obtain
algebraic corrections to the linear growth law of $\delta\bar
R_\parallel(t)$, because $\pd_\xi\varphi_1(0)=0$. Higher-order terms
in the expansion (as far as they are analytically tractable) turn out
to be ill-behaved near the ends.

\paragraph*{Homogeneous relaxation ($\tlpar\ll t\ll \tR$).}
The subsequent regime of homogeneous tension relaxation is exactly
equivalent to the one of the ``force'' case (see the respective
boundary conditions Eqs.~(\ref{eq:bc-force}, \ref{eq:bc-shear})), and
the results
Eqs.~(\ref{eq:solution-homogeneous-relaxation-force-g}, \ref{eq:solution-homogeneous-relaxation-force-h-characteristics})
apply here as well.

\subsubsection{\label{sec:quench}``Quench'' setup}

This scenario, with the initial and boundary conditions
\begin{equation}\label{eq:bc-quench}
  \begin{aligned}
    \bar f(s,t < 0)&=0 \\
    \lp(t < 0)&=\theta\lp
  \end{aligned}
  \quad \text{and }\quad
  \begin{aligned}
    \bar f(0,t>0)&=\bar f(L,t>0)=0\\
    \lp(t > 0)&=\lp
  \end{aligned}
\end{equation}
has been introduced as ``$\lp$-quench'' in
Ref.~\cite{hallatschek-frey-kroy:05}. In contrast to the scenarios
discussed above, we lack a quantity providing a force scale, and the
tension attains a simple scaling form in the propagation and
relaxation regimes, i.e., there is no linear--nonlinear crossover.
However, this scaling form still strongly depends on the value of
$\theta$. Physically the limit $\theta\to 0$ corresponds to suddenly
switching off thermal forces for a thermally equilibrated filament,
hence purely deterministic relaxation (see
Ref.~\cite{hallatschek-frey-kroy:04}). A small quench $\theta\approx
1$ will not induce strong tension, but the limit $\theta\to\infty$
describes the scenario of a completely straight contour that
equilibrates purely under the action of stochastic forces, and the
resulting tension may be very large at short times.  In this case,
similar approximations as employed when discussing the nonlinear
regime in Sec.~\ref{sec:force} can be justified. Assuming $\bar
F^2/t\gg 1$, the response function $\chi_\perp(q;t,t')$ from
Eq.~\eqref{eq:chi-perp} is finite only near $t'\approx t$ which
suggests the linearization $\bar F(s,t)-\bar F(s,t')\approx [\pd_t
\bar F(s,t)](t-t')$ in the exponent.  Performing the $t'$-integral in
the second term of Eq.~\eqref{eq:pide} yields
\begin{equation*}
  \pd_s^2 \bar F \approx \hat\zeta \int_0^\infty\!\frac{\td q}{\pi\lp}\left[
    \frac{1-\chi_\perp^2(q;t,0)}{\theta q^2} - 
    \frac{1-\chi_\perp^2(q;t,0)}{q^2 + \pd_t \bar F}\right].
\end{equation*}
In contrast to Eq.~\eqref{eq:pide-qs}, we may not set $\chi_\perp\to
0$ in the first term, because this would produce an IR-divergence. But
we can neglect the bending contribution $q^4 t$ in the exponent of the
first and set $\chi_\perp\to 0$ only in the second term. This gives
\begin{align}
  \pd_s^2 \bar F &\approx \hat\zeta\int_0^\infty\!\frac{\td
    q}{\pi\lp}\left[ \frac{1-\e^{-2 q^2 \bar F}}{\theta q^2} -
    \frac{1}{q^2 + \pd_t \bar F}\right]
  \nonumber\\
  \label{eq:pide-nonlinear-quench}
  &=
  \frac{\hat\zeta}{2\lp}\left[\tfrac{2}{\theta}\sqrt{\tfrac{2}{\pi}\bar
      F} - (\pd_t \bar F)^{-1/2}\right].
\end{align}

\paragraph*{Propagation ($t\ll\tlpar$).} Inserting the scaling ansatz
$\bar F(s,t)=\theta t^{1/2} \phi(s/\lpar(t))$ with $\lpar(t) =
(\lp/\hat\zeta)^{1/2} \theta^{3/4} t^{1/8}$ removes the parameter
dependence in Eq.~\eqref{eq:pide-nonlinear-quench}:
\begin{equation}\label{eq:ode-nonlinear-quench}
  \pd_\xi^2 \phi(\xi)= \sqrt{\tfrac{2}{\pi}\phi(\xi)} - 
  \left[2\phi(\xi)-\tfrac{1}{2}\xi\pd_\xi\phi(\xi)\right]^{-1/2}.
\end{equation}
Boundary conditions are $\phi(0)=0$ and $\pd_\xi\phi(\xi\to\infty)=0$.
From a numerical solution we obtain $\phi(\xi\to\infty)=\sqrt{\pi}/2$
as expected from Eq.~\eqref{eq:ode-nonlinear-quench} and
$\phi'(0)\approx 1.44$.  The assumption $\bar F^2/t \simeq \theta \gg
1$ is justified for all times $t\ll\tlpar = L^8/(\lp^4\theta^6)$.

\paragraph*{Homogeneous relaxation ($\tlpar\ll t\ll \tR$).}
Because we expect universal long time relaxation in the strong
quenching limit $\theta\to\infty$ similar to the ``force''-case, we
try the ansatz $\bar F(s,t)= t^{1/3} (\hat\zeta
L^2/\lp)^{2/3}\phi(\xi)$ with $\xi=s/L$ in
Eq.~\eqref{eq:pide-nonlinear-quench}:
\begin{equation}\label{eq:pde-nonlinear-quench}
  \begin{split}
    \pd_\xi^2 \phi &= \left(\frac{\hat\zeta
        L^2}{\lp}\right)^{2/3}\theta^{-1}
    t^{-1/6}\sqrt{\frac{2}{\pi}\phi}-\sqrt{\frac{3}{4\phi}} \\
    &\approx -\sqrt{\frac{3}{4\phi}} \qquad\text{if }t\gg
    \frac{L^8}{\lp^4\theta^6}=\tlpar.
  \end{split}
\end{equation}
Because now $\bar f(s,t)=\pd_t \bar F(s,t)=g(t) h(\xi)$ with
$g(t)=(\hat\zeta L^2/\lp t)^{2/3}$ and $h(\xi)=\phi(\xi)/3$ as in
Eqs.~(\ref{eq:solution-homogeneous-relaxation-force-g},
\ref{eq:solution-homogeneous-relaxation-force-h}), this regime of
homogeneous relaxation is identical to the one in the
``force''-case. The condition $\bar F^2/t\gg 1$ holds until $t\simeq
t_\star=L^8/\lp^4=\theta^6\tlpar$, which is usually already larger
than $\tR$ if $\lp\ll L$.

\subsection{\label{sec:quantitative-observables}Results for pertinent observables}
The maximum bulk tension $\fb(t)=\bar f(L/2,t)$ (in the ``field''
case, we prefer to use the grafting force $\fg(t)=f(0,t)$) can be
obtained directly from the tension profiles computed in the preceding
section. The change in end-to-end distance $\delta R_\parallel(t)$
follows from a simple formula: With the sign convention used before,
this change has to equal the total amount of stored length that has
been created.  Hence, we integrate the ensemble averaged change in
stored length density $\pd_t \ave{\varrho}(s,t)$ over $s$ and $t$:
\begin{equation}\label{eq:deltar-1}
  \delta R_\parallel(t) = \int_0^L\!\td s\int_0^t\!\td
  t'\pd_t\ave{\varrho}(s,t').
\end{equation}
Defining $\ave{\varrho}=\ave{\bar\varrho}+\ave{\varrho^\text{e}}$ in
Eq.~\eqref{eq:rho} from the decomposition
Eq.~\eqref{eq:mode-decomposition}, we obtain $\delta
R_\parallel(t)=\delta \bar R_\parallel(t)+\delta
R^\text{e}_\parallel(t)$. The first part obeys the deterministic
Eq.~\eqref{eq:eom-mspt-L}:
\begin{equation}\label{eq:deltar-bulk}
  \begin{split}
    \delta \bar R_\parallel(t)  &= -\hat\zeta^{-1} \int_0^L\!\td s\int_0^t\!\td t' \pd_s^2 f(s,t') \\
    &= -\hat\zeta^{-1} \left[ \pd_s \bar F(L,t) - \pd_s \bar
      F(0,t)\right].
  \end{split}
\end{equation}
It accounts only for the ``slow'' coarse-grained tension dynamics but neglects
subdominant and stochastic contributions $\delta
R^\text{e}_\parallel(t)$ from ``fast'' fluctuating boundary segments
analyzed in the next section.  Results for $\delta \bar
R_\parallel(t)$ and $\fb(t) (\fg(t))$ are summarized in
Tables~\ref{tb:deltar-scaling} and \ref{tb:bulk-tension-scaling}.

\begin{table}
  \begin{tabular}{>{$}c<{$}|>{$}c<{$}|>{$}c<{$}|>{$}c<{$}}
    \hline
    \delta\bar R_\parallel(t) & t \ll \tf &  
    \tf\ll t \ll \tlpar & \tlpar \ll t \\
    \hline\hline
    \text{``force''}&
    \frac{2^{5/8}\fpre t^{7/8}}{\EGamma{15/8}\hat\zeta^{1/2}\lp^{1/2}} & 
    2.48 \frac{\fpre^{1/4} t^{1/2}}{\hat\zeta^{1/2}\lp^{1/2}} &
    \left(\frac{18 L t}{\hat\zeta\lp^2}\right)^{1/3} \\
    \hline
    \text{``field''} &
    \multicolumn{2}{c|}{$ g t/\hat\zeta$} & 
    \left(\frac{9 L t}{2 \hat\zeta\lp^2}\right)^{1/3} \\
    \hline
    \text{``shear''} &
    \dot\gamma t L\,\left[1-\mathcal{O}(\lpar/L)\right] &
    \dot\gamma t L &
    \left(\frac{18 L t}{\hat\zeta\lp^2}\right)^{1/3} \\
    \hline
    \text{``quench''} &
    \multicolumn{2}{c|}{$2.88 \frac{\theta^{3/4} t^{3/8}}{\hat\zeta^{1/2}\lp^{1/2}}$} & 
    \left(\frac{18 L t}{\hat\zeta\lp^2}\right)^{1/3} \\
    \hline\hline
  \end{tabular}
  \caption{\label{tb:deltar-scaling}Asymptotic scaling laws for the
    change in projected length $\delta\bar R_\parallel(t)$ from
    Eq.~\eqref{eq:deltar-bulk} for the different setups. Units have been chosen
    such that $\kappa\equiv\zeta_\perp\equiv 1$.}
\end{table}
\begin{table}
  \begin{tabular}{>{$}c<{$}|>{$}c<{$}|>{$}c<{$}|>{$}c<{$}}
    \hline
    \fb(t),\fg(t) & t \ll \tf &  
    \tf\ll t \ll \tlpar & \tlpar \ll t \\
    \hline\hline
    \text{``force''}&
    \multicolumn{2}{c|}{$\fpre$} &
    \left(\frac{3 \hat\zeta L^2}{128 \lp t}\right)^{2/3} \\
    \hline
    \text{``field''} &
    \multicolumn{2}{c|}{$g L [1- \mathcal{O}(\lpar/L)]$} &
    \left(\frac{3 \hat\zeta L^2}{32 \lp t}\right)^{2/3} \\
    \hline
    \text{``shear''} &
    \frac{1}{8}\hat\zeta\dot\gamma L^2 &
    \frac{1}{8}\hat\zeta\dot\gamma L^2[1 - \mathcal{O}(t/\tlpar)] &
    \left(\frac{3 \hat\zeta L^2}{128 \lp t}\right)^{2/3} \\
    \hline
    \text{``quench''} &
    \multicolumn{2}{c|}{$\tfrac{1}{4}\pi^{1/2}\theta\, t^{-1/2}$} &
    \left(\frac{3 \hat\zeta L^2}{128 \lp t}\right)^{2/3} \\
    \hline\hline
  \end{tabular}
  \caption{\label{tb:bulk-tension-scaling}Asymptotic scaling laws
    for the maximum bulk tension $\fb(t)=\bar f(L/2,t)$ (grafting force
    $\fg(t)=\bar f(0,t)$) for different setups. Units as in Table~\ref{tb:deltar-scaling}.}
\end{table}

\subsection{Boundary effects}
Because our theory applies to times $t\ll \tR$ long before the
relaxation of long-wavelength modes becomes relevant, because
Eq.~\eqref{eq:tension-dynamics} results from a coarse-grained
description that averages over small-scale fluctuations, and finally
because projecting the end-to-end distance onto the longitudinal axis
suppresses some end effects~\cite{hallatschek-frey-kroy:07b}, the
dependence of $\delta R_\parallel=\delta \bar R_\parallel+\delta
R^\text{e}_\parallel$ on the boundary conditions for the contour
$\bvec r(s)$ is only subdominant but still non-negligible.  While the
``bulk contribution'' $\delta \bar R_\parallel$ is independent of
boundary effects and dynamically self-averaging, this stochastic
dependence is accounted for by an additional ``end contribution''
$\delta R^\text{e}_\parallel$, which stems from the oscillating term
$c_q(s)$ in Eq.~\eqref{eq:mode-decomposition}, and decays rapidly on
much smaller length scales than that of tension
variations~\cite{hallatschek-frey-kroy:07a}. We may therefore evaluate
this part at the boundaries under zero tension (i.e., using
$\chi_\perp(q;t,t')=\e^{-2 q^4 (t-t')}$ instead of
Eq.~\eqref{eq:chi-perp}):
\begin{multline}\label{eq:deltar-e}
  \delta R_\parallel^\text{e}(t) \approx -\int_0^L\!\td s
  \int_0^\infty\!
  \frac{\td q}{\pi\lp} \frac{q^2-\theta( q^2+\bar f_0(s))}{q^2\theta(q^2+\bar f_0(s))}\\
  \times (1-\e^{-2 q^4 t}) c_q(s).
\end{multline}
Consistent with this simplification, we approximate the $w_q(s)$ by
eigenfunctions of the biharmonic operator $\pd_s^4$ (see
Ref.~\cite{wiggins-etal:98}). Again exploiting the scale separation in
this integral over the rapidly fluctuating term $c_q(s)$, we use only
the spatial average $\bar\fb$ of the slowly varying prestress $\bar
f_0(s)$. This means $\fpre\to\bar\fb=\frac{1}{12}\hat\zeta\dot\gamma
L^2$ for the ``shear'' case, and $\fpre\to\bar\fb=\frac{1}{2}g L$ for
the ``field'' setup. Because there is only one free end in the latter,
the contribution to $\delta R^\text{e}_\parallel$ is one half of the
``force'' result.

\paragraph*{Free ends.} 
If $w_q''=w_q'''=0$ at $s=0,L$, we use
\begin{multline}\label{eq:wq-free}
  w_q(s)=\frac{1}{\sqrt{L}}\Big[\frac{\sin q L+\sinh qL}{\cos qL-\cosh
    qL} (\sin q s+\sinh q s)\\+\cos q s + \cosh q s\Big],
\end{multline}
where $q$ is a solution of $\cos q L\cosh q L=1$. For $t\ll\tR$, the
$q$-integral in Eq.~\eqref{eq:deltar-e} is dominated by short
wavelength contributions, and for the relevant asymptotically large
modes the $s$-integral over $c_q(s)$ reads
\begin{equation}
  \int_0^L\!\td s\, c_q (s) = \frac{6}{q} + \mathcal{O}(\e^{-qL}).
\end{equation}

\paragraph*{Hinged ends.}
For $w_q=w_q''=0$ at $s=0,L$, we obtain
\begin{equation}
  w_q(s)=\sqrt{\tfrac{2}{L}} \sin q s
\end{equation}
with $\sin q L=0$. In this case, $\int_0^L\!\td s\, c_q(s)=0$.

\paragraph*{Clamped ends.}
Here, $w_q=w_q'=0$ at $s=0,L$, and we have
\begin{multline}
  w_q(s)= \frac{1}{\sqrt{L}}\Big[\sin q s-\sinh q s +  \\
  \frac{\cos q L - \cosh q L}{\sin q L+\sinh q L}(\cos q s - \cosh q
  s)\Big],
\end{multline}
with $\cos q L\cosh q L=1$ and $\int_0^L\!\td s\, c_q(s)\sim - 2/q$.
Up to a prefactor, the contributions for clamped ends is identical to
the one for free ends.

\paragraph*{Torqued ends.}
If $w_q'=w_q'''=0$ at $s=0,L$, the Eigenmodes are
\begin{equation}
  w_q(s)=\sqrt{\tfrac{2}{L}} \cos q s
\end{equation}
with $\sin q L=0$ and $\int_0^L\!\td s\, c_q(s)=0$.

Using the asymptotic limit of $\int_0^L c_q(s)\td s$, we evaluate the
end contributions in the free ends situation of our setups:

\paragraph*{``Force'' setup.} Here, we obtain
\begin{equation}
  \delta R^\text{e}_\parallel = \frac{6\fpre t}{\pi\lp} G(\fpre t^{1/2}),
\end{equation}
where
\begin{equation}
  \begin{split}
    G(\phi) =& \int_0^\infty\!\td k \frac{1-\e^{-2 k^4}}{k^3 (k^2 + \phi)}  \\
    =& -\frac{1}{4\phi^2}\Big[\e^{-2\phi^2}(\pi\erfi\sqrt{2\phi^2} - \Ei (2\phi^2))\\
    & + \ln2\phi^2 + \gamma_\text{E} - \sqrt{8\pi\phi^2}\Big],
  \end{split}
\end{equation}
with $\erfi$ the imaginary error function, $\Ei$ the exponential
integral and $\gamma_\text{E}\approx 0.577$ Euler's constant.  The
asymptotic behavior is
\begin{subequations}
  \begin{empheq}[left={\displaystyle{\delta R_\parallel^\text{e}(t)
        \sim \empheqlbrace}}]{align} &-\frac{3\fpre t}{\pi\lp}\ln 2
    \fpre^2 t,
    & t\ll\fpre^{-2}\\
    &\frac{6 t^{1/2}}{\sqrt{2\pi}\lp},& t\gg\fpre^{-2}.
  \end{empheq}
\end{subequations}

\paragraph*{``Quench'' setup.}
For free ends, we obtain from Eq.~\eqref{eq:deltar-e}:
\begin{equation}
  \begin{split}
    \delta R_\parallel^\text{e} &= \frac{6 t^{1/2}}{\pi\lp}
    \left(1-\frac{1}{\theta}\right)\int_0^\infty\!\td k
    \frac{1-\e^{-2 k^4}}{k^3}\\
    &=\sqrt{\frac{18}{\pi}}\frac{t^{1/2}}{\lp}\left(1-\frac{1}{\theta}\right).
  \end{split}
\end{equation}

As anticipated~\cite{hallatschek-frey-kroy:07b}, the end contribution
$\delta R^\text{e}_\parallel$ is zero for hinged and torqued ends,
while it leads to an additional reduction of $R_\parallel$ for free
ends and has a lengthening effect for clamped ends. Note that indeed
$\delta R_\parallel^\text{e}\ll\delta\bar R_\parallel$ is only
subdominant for $t\ll \tR$, but the quantitative relevance of this
contribution becomes evident when it is directly compared to numerical
solutions of Eq.~\eqref{eq:pide} and simulation data in non-asymptotic
regimes.

\section{\label{sec:comparison}Comparison to simulation data}

\begin{figure*}
\begin{minipage}{.48\textwidth}
  \psfrag{X1}{$\tau/\tau_0$} \psfrag{X2}{$\tau/\tau_0$}
  \psfrag{Y1}{$\delta R_\parallel(t)/L$} \psfrag{Y2}{$\bar \fb(t)/(\kb
    T/b)$} \psfrag{A}{(a)}
  \includegraphics[angle=270,width=\textwidth]{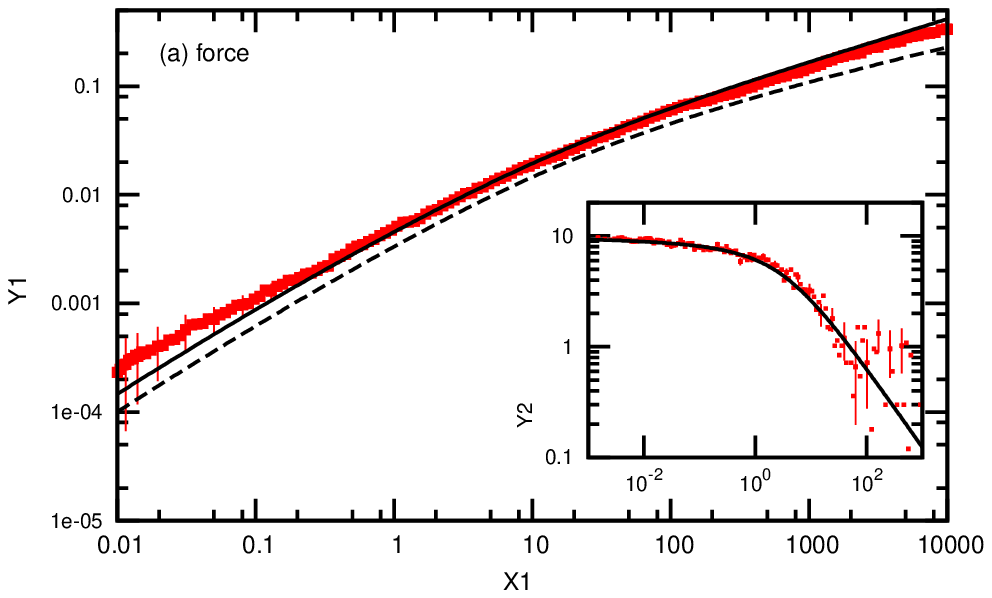}
\end{minipage}\hfill
\begin{minipage}{.48\textwidth}
  \psfrag{X1}{$\tau/\tau_0$} \psfrag{X2}{$\tau/\tau_0$}
  \psfrag{Y1}{$\delta R_\parallel(t)/L$} \psfrag{Y2}{$\bar \fb(t)/(\kb
    T/b)$}\psfrag{A}{(b)}
  \includegraphics[angle=270,width=\textwidth]{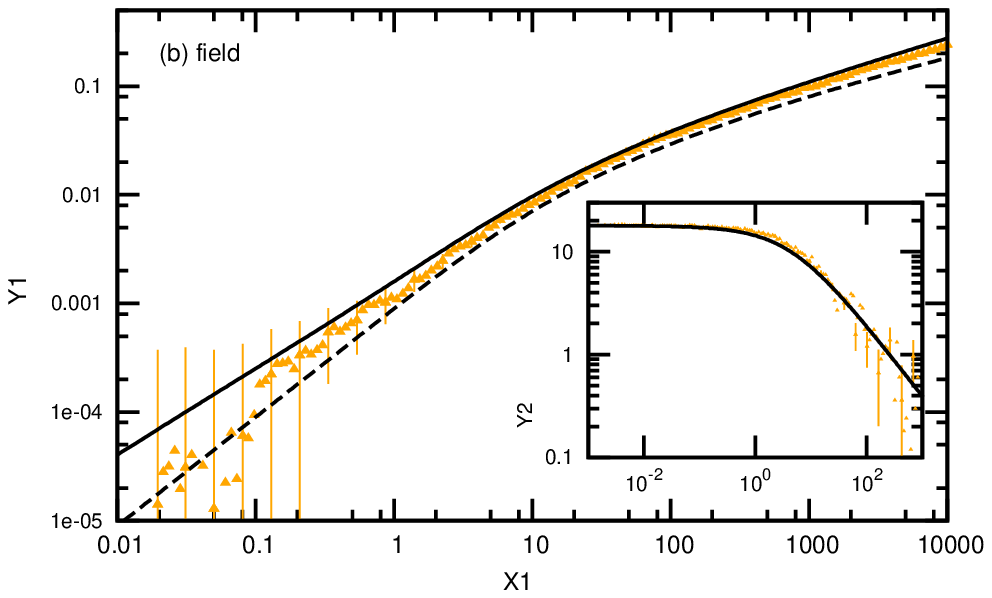}
\end{minipage}

\begin{minipage}{.48\textwidth}
  \psfrag{X1}{$\tau/\tau_0$} \psfrag{X2}{$\tau/\tau_0$}
  \psfrag{Y1}{$\delta R_\parallel(t)/L$} \psfrag{Y2}{$\bar \fb(t)/(\kb
    T/b)$}\psfrag{A}{(c)}
  \includegraphics[angle=270,width=\textwidth]{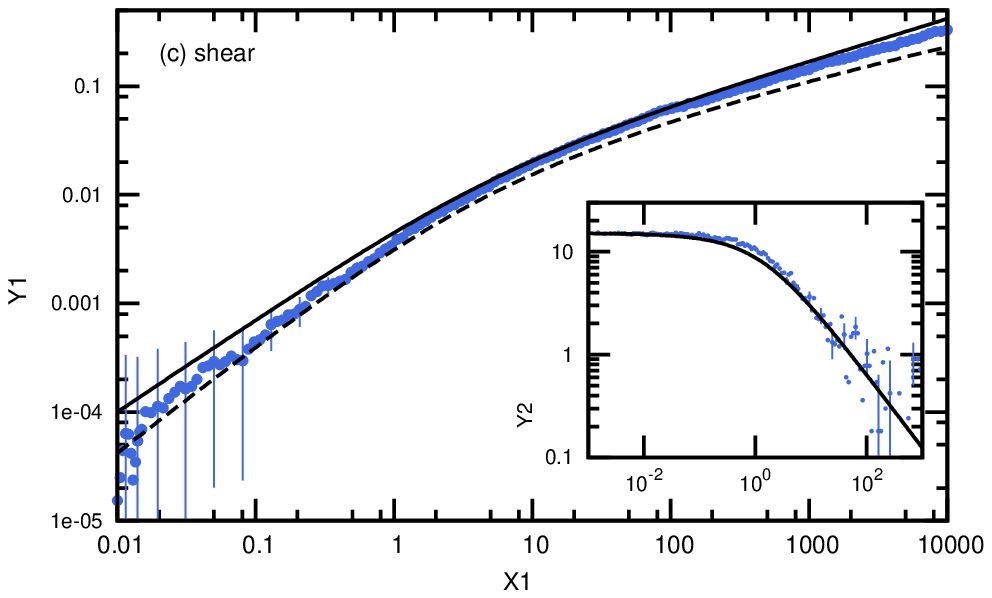}
\end{minipage}\hfill
\begin{minipage}{.48\textwidth}
  \psfrag{X1}{$\tau/\tau_0$} \psfrag{X2}{$\tau/\tau_0$}
  \psfrag{Y1}{$\delta R_\parallel(t)/L$} \psfrag{Y2}{$\bar \fb(t)/(\kb
    T/b)$}\psfrag{A}{(d)}
  \includegraphics[angle=270,width=\textwidth]{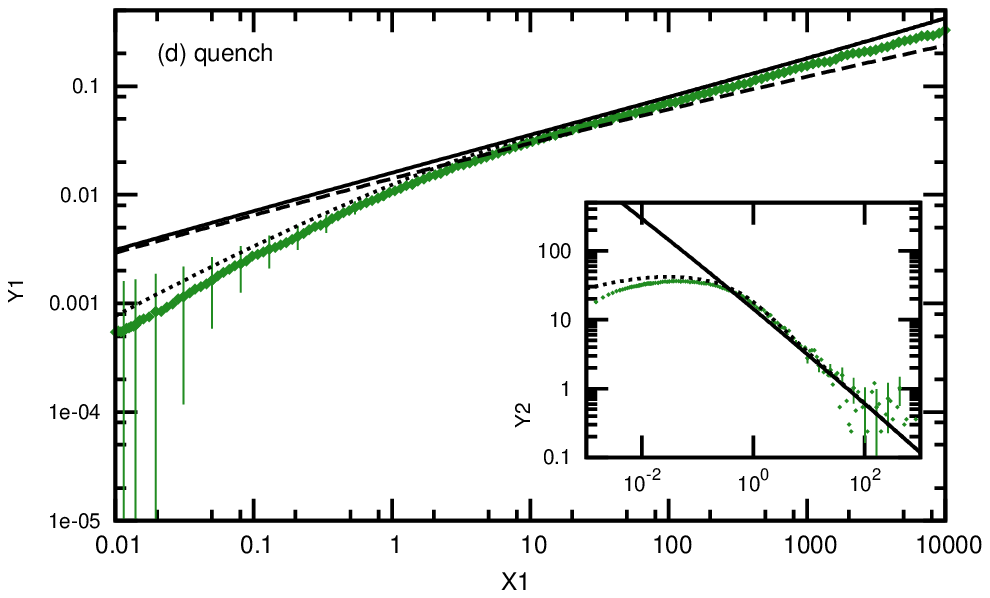}
\end{minipage}
\caption{\label{fig:comparison}(Color online) Comparison between
  theory and simulation data for a ``force'' (a), ``field'' (b),
  ``shear'' (c), and ``quench'' (d) scenario, respectively, for
  $\delta R_\parallel(t)$ and $\bar \fb(t)$ (insets). Note that the
  bulk contribution $\delta \bar R_\parallel(t)$ from
  Eq.~\eqref{eq:deltar-bulk} (dashed lines) underestimates the
  contraction, which is corrected for by including the end
  contribution $\delta R^\text{e}_\parallel(t)$ from
  Eq.~\eqref{eq:deltar-e} (solid lines). In the simulation of the
  ``quench'' scenario, the tension follows the sudden change in
  temperature only with a delay related to the longitudinal
  propagation of backbone strain, which leads via
  Eq.~\eqref{eq:observables} to a different scaling of $\delta
  R_\parallel$ at short times. This can be accounted for by a
  correction term (dotted line) including the finite backbone
  extensibility of the bead-spring model~\cite{obermayer:unpub}.
  Simulation data as in Fig.~\ref{fig:quantitative-results}.}
\end{figure*}

The asymptotic scaling laws of Fig.~\ref{fig:qualitative-results} are
derived in the limit $\fpre\to\infty$, and the difference to the
numerical solutions gets smaller than 20\% only for $\fpre \gtrsim
10^{10}\kb T \lp^3/L^4$. While this can easily be realized in
experiments, for instance on DNA (cf.\ Table~\ref{tb:numbers} below),
it is not possible in simulations due to the usual trade-off between
computational efficiency and accuracy. For a comparison between our
theoretical results and the simulation data of
Fig.~\ref{fig:quantitative-results}, we therefore compute the bulk
part $\delta \bar R_\parallel(t)$ and $\bar \fb(t)$ using numerical
solutions to Eq.~\eqref{eq:pide} as described
previously~\cite{obermayer-hallatschek-frey-kroy:07}.
Fig.~\ref{fig:comparison} shows simulation data and analytical results
for all four scenarios, using $\mu=(\zeta_\perp b)^{-1}$ and
$\zeta_\perp=\zeta_\parallel$ to relate the (isotropic) mobility of
the beads in the simulation model to the anisotropic friction
coefficients per length of the continuous wormlike chain used in our
theory.  The bulk contribution $\delta\bar R_\parallel(t)$ (dashed
lines), while having the correct qualitative behavior, underestimates
the contraction by as much as 50\%. Including end fluctuations with
$\delta R^\text{e}_\parallel(t)$ (solid lines) gives results for
$\delta R_\parallel(t)$ that are slightly overestimated for longer
times.  This could be caused by possibly oversimplifying
approximations made when evaluating Eq.~\eqref{eq:deltar-e}, or by a
gradual breakdown of the weakly-bending limit.

In the ``quench'' case, we observe a strong deviation between
simulation and theory for short times, both in $\delta R_\parallel$
and $\bar\fb$. While $\bar\fb\propto t^{-1/2}$ diverges as $t\to 0$ in
our theory, the actual tension in the simulations is finite. This is
due to the extensible backbone of a bead-spring chain: the tension
follows the sudden change in environmental conditions only with a
temporal delay related to the finite propagation speed of longitudinal
backbone strain. Because now $\bar\fb$ is smaller than predicted, the
contraction $\delta R_\parallel(t)$ is also reduced, see the scaling
relation Eq.~\eqref{eq:observables}.  It is, however, possible to
include a finite extensibility correction in
Eq.~\eqref{eq:tension-dynamics}. Because this nontrivial extension is
only marginally relevant for the present discussion, which is focused
on differences in the relaxation dynamics from an initially straight
conformation, we present a detailed discussion
elsewhere~\cite{obermayer:unpub}, and merely show the corrected
results for $\delta R_\parallel(t)$ and $\bar\fb(t)$ for the
``quench'' case in Fig.~\ref{fig:comparison}(d) (dotted lines).  The
analysis of this correction term allowed to choose parameters such
that our results are not affected by microscopic details for the other
setups, but this was not possible for the ``quench'' case with its
singular short-time behavior.

Altogether, we now obtain good quantitative agreement between computer
simulation and theory for all four setups and both observables over
six decades in time \emph{without adjustable parameters}. Having
reliable theoretical control over the relaxation dynamics, we will now
present quantitative estimates for the feasible choice of control
parameters in experiments and a qualitative discussion of the
influence of some additional important effects.

\section{Experimental implications}

\begin{table}
  \centering
  \begin{tabular*}{\hsize}{@{\extracolsep{\fill}}cc}
    \begin{tabular*}{.45\hsize}{@{\extracolsep{\fill}}ccc}
      (a) &  DNA & actin \\
      \hline
      $\tf$ & $10^{-7}\s$ & $10^{-5}\s$ \\ 
      $\tlpar$ & $0.05\s$ & $0.003\s$ \\
      $\tR$ & $\approx 6\s$ & $\approx 10\s$ \\
      \hline
    \end{tabular*} &
    \begin{tabular*}{.45\hsize}{@{\extracolsep{\fill}}ccc}
      (b) & DNA  &  actin  \\
      \hline
      $\fc$ & $0.08\pN$ & $0.2\,\mathrm{fN}$ \\
      $\vphantom{\tlpar}v_\text{c}$ & $27\mum/\s$&$72\nm/\s$\\
      $\dot\gamma_\text{c}$ & $6.6\s^{-1}$ & $0.02\s^{-1}$ \\
      \hline
    \end{tabular*}
  \end{tabular*}
  \caption{\label{tb:numbers} (a) Characteristic time scales (given a force 
    of $\fpre=2\pN$) and (b) bounds on control parameters for typical DNA~\cite{bohbot_raviv-etal:04} 
    ($L\approx20\mum$, $\lp\approx50\nm$) and actin~\cite{legoff-hallatschek-frey-amblard:02} ($L\approx20\mum$, 
    $\lp\approx17\mum$) in solution with viscosity 
    $\eta\approx 10^{-3}\Pa\,\s$ at room temperature.}
\end{table}

\subsection{Time and force scales}
In Table~\ref{tb:numbers} we have compiled numerical examples for the
various time and force scales introduced above based on literature
values for DNA~\cite{bohbot_raviv-etal:04} and
actin~\cite{legoff-hallatschek-frey-amblard:02}.  In order to obtain
sufficiently straight initial conformations, a conservative estimate
for the control parameters $\fpre$, $v$, and $\dot\gamma$ requires
them to be chosen by a factor 25 larger than the respective values
$\fc$, $v_\text{c}$, and $\dot\gamma_c$. The quenching strength
$\theta$ should be significantly larger than $\theta_\text{c}= L/\lp$.
The crossover times $\tlpar\propto L^2/\fpre^{3/2}$ and
$\tf\propto\fpre^{-2}$ depend strongly on the adjustable quantities
$L$ and $\fpre$; hence the time window of interest can be varied
considerably between scenario-specific and universal relaxation.  The
algebraic relaxation ends at times near $\tR$, for which we can give
only a rough estimate as the unknown numerical prefactor is influenced
by boundary conditions and hydrodynamic interactions and may
substantially differ from
unity~\cite{legoff-hallatschek-frey-amblard:02}.

\subsection{Onset of the stretch--coil transition.}  
Since experiments are often performed using quite flexible polymers
like DNA with $\lp\ll L$, the weakly-bending approximation will
finally become invalid in regions near the ends, where the contour
starts to (literally) coil up as the tension relaxes.  Borrowing ideas
from flexible polymer theory allows to derive scaling laws accounting
for the onset of the stretch--coil transition.  The stem--flower
picture of Brochard-Wyart~\cite{brochard:95} describes transient
relaxation processes of flexible polymers with Kuhn length $a$ and
friction coefficient $\zeta$.  Entropic forces on the order of $\kb
T/a$ arising in the bulk pull the ends inwards. Balancing these forces
with the associated friction gives the well-known scaling $\ell_\ast
\simeq [\kb T t/\zeta a]^{1/2}$ for the growth of ``flowers'' leading
to an additional longitudinal contraction.  

In the case of strongly stretched semiflexible polymers, this
correction is negligible on time scales $t\ll\tR$. Here, the Kuhn
segments are of size $\lp$, and their Rouse-like relaxation after
internal bending modes have equilibrated would generate flowers of
size $\ell_\ast\simeq L (t/\tR)^{1/2}$.  However, in the relevant
universal regime of homogeneous tension relaxation ($\tlpar\ll t \ll
\tR$), the bulk tension $\fb\simeq \kb T (\tR/t)^{2/3}/\lp$ (see
Table~\ref{tb:bulk-tension-scaling}) is much larger than $\kb T/\lp$,
and the ends are pulled inwards so fast that a flower of the above
size would be too large for the resulting drag.  Observing that the
associated roughly parabolic tension
profiles~\cite{hallatschek-frey-kroy:07b} attain values of about $\kb
T/\lp$ within distances $\ell'_\ast \simeq L(t/\tR)^{2/3}$ from the
ends, one easily confirms that this smaller value for the flower's
size indeed restores the friction balance.  Altogether, we find that
for times $t\ll\tR$ the $(t/\tR)^{2/3}$-growth of ``flower''-like
end-regions, where the weakly-bending approximation breaks down, is
subdominant against the $(t/\tR)^{1/3}$-contraction of the remaining
weakly-bending part of the filament. Only at $t\simeq\tR$, our
assumptions finally cease to hold and more appropriate models, for
conformational relaxation as well as hydrodynamic interactions, need
to be employed (see, e.g., Ref.~\cite{hoffmann-shaqfeh:07}).  In our
simulation, we do not expect a pronounced stretch--coil transition
because the number $L/\lp$ is not large enough. We also checked that
global rotational diffusion~\cite{doi-edwards:86}, apparently reducing
the longitudinal projection of the end-to-end distance, can be
neglected.

\subsection{Hydrodynamic interactions}
Finally, we want to briefly comment on hydrodynamic interactions.
Their pronounced effects for strongly coiled polymers reduce to mere
logarithmic corrections for relatively straight
filaments~\cite{doi-edwards:86}. As suggested
previously~\cite{bohbot_raviv-etal:04}, we speculate that these
corrections can be summarily included via a phenomenological
renormalization $L\to\ell_\text{eff}$ in the friction coefficient
$\zeta\propto\eta/\ln(L/b)$~\cite{doi-edwards:86}, where $\eta$ is the
solvent's viscosity and $b$ the monomer size. Within our model, this
has almost no further consequences than slightly shifting the time
unit, cf.~Eqs.~\eqref{eq:tension-dynamics} and \eqref{eq:eom-mspt}.
An appropriate time rescaling compensates for changes in the friction
coefficients, and could therefore easily be checked in experiments.
The setups of Refs.~\cite{goshen-etal:05,crut-etal:07}, where
initially stretched DNA relaxes with one end attached to a wall and
the other fixed to a bead, can easily be modeled within our theory by
appropriately adjusting the boundary conditions for the tension. It
turns out that spatial inhomogeneities of the tension and end
fluctuations are suppressed and hydrodynamic interactions (primarily
between bead and wall) are enhanced, such that a simple
quasi-stationary approach describes the data very well.

\section{Conclusion}

We have presented a comprehensive theoretical analysis of the
conformational relaxation dynamics of semiflexible polymers from an
initially straight conformation.  Special emphasis has been put on a
systematic investigation of four fundamentally different realizations
of ``initially straight''.  The sudden removal of the straightening
constraint leads in all cases to strong spatial inhomogeneities of the
filaments' backbone tension. Analyzing two exemplary and easily
accessible observables, we found that for short times, when these
nontrivial spatial variations are restricted to the boundaries, the
relaxation dynamics crucially depends on the actual initial
conditions: polymers pre-stretched with forces display tension
propagation effects, in contrast to chains straightened by fields or
flows, and a quench leads to yet other effects. In the universal
relaxation regime at longer times, the tension becomes
quasi-statically equilibrated and independent of initial conditions,
but its spatial inhomogeneity remains relevant. Additionally to the
derivation of asymptotic growth laws, we extended the systematic
theory of
Refs.~\cite{hallatschek-frey-kroy:05,hallatschek-frey-kroy:07a,hallatschek-frey-kroy:07b}
to include the surprisingly important influence of different boundary
conditions. For non-asymptotic parameter values, quantitative and
parameter-free agreement between simulation data and theory could be
achieved over six time decades below the filament's longest relaxation
time. In the ``quench'' case, short-time deviations could be
attributed to the finite backbone extensibility of the bead-spring
chains used in the simulations. Finally, we discussed quantitative
implications for possible experimental realizations, adapted a widely
used scaling argument for the onset of the stretch--coil transition
for flexible polymers to the semiflexible case (dominated by bending
energy), and commented on hydrodynamic interactions.  We hope that our
thorough discussion of the non-equilibrium dynamics of an initially
straight polymer will help to design new quantitative single molecule
experiments and lead to a better understanding of more complex
phenomena such as force transduction and recoil of disrupted stress
fibers in cells~\cite{kumar-etal:06}.

\begin{acknowledgments}
  We gratefully acknowledge financial support via the German Academic
  Exchange Program (DAAD) (OH), by the Deutsche Forschungsgemeinschaft
  (DFG) through grant no. Ha 5163/1 (OH), SFB 486 (BO, EF), and FOR
  877 (KR 1138/21-1) (KK), and of the German Excellence Initiative via
  the programs ``Nanosystems Initiative Munich (NIM)'' (BO, EF) and
  Leipzig School of Natural Sciences ``Building with molecules and
  nano-objects'' (KK).
\end{acknowledgments}


\end{document}